\documentclass[12pt]{article}
\usepackage[top=2cm,bottom=3cm,left=2cm,right=2cm]{geometry}

\usepackage[utf8]{inputenc}
\usepackage{amsmath,amssymb,amsthm}
\usepackage{simplewick}
\usepackage{graphicx}
\usepackage{cancel}
\usepackage[unicode, pdfencoding=auto, psdextra, linktocpage]{hyperref}
\usepackage{comment}
\usepackage{cite}

\begin{document}

\begin{titlepage}
\rightline\today

\begin{center}

\vspace{3.5cm}

{\large \bf{A two parameter family of lightcone-like hyperbolic string vertices}}

\vspace{1cm}

{\large Vinícius Bernardes$^{1}$ and Ulisses Portugal$^{1,2}$}

\vspace{1cm}

{\large $^1$} \!\!\!\! {\it CEICO, FZU - Institute of Physics}\\
{\it of the Czech Academy of Sciences}\\
{\it Na Slovance 2, 182 21 Prague 8, Czech Republic}\\

\vspace{.5cm}

{\large $^2$} \!\!\!\! {\it  ICTP South American Institute for Fundamental Research}\\
{\it Instituto de Física Teórica, UNESP - Univ. Estadual Paulista}\\
{\it Rua Dr. Bento Teobaldo Ferraz 271, 01140-070, São Paulo, SP, Brazil}\\

\vspace{1cm}

\end{center}

\begin{abstract}
    We introduce a two parameter family of string field theory vertices, which we refer to as hyperbolic Kaku vertices. It is defined in terms of hyperbolic metrics on the Riemann surface, but the geometry is allowed to depend on inputs of the states. The vertices are defined for both open and closed strings. In either case, the family contains the hyperbolic vertices. Then we show that the open string lightcone vertex is obtained as the flat limit of the hyperbolic Kaku vertices. The open string Kaku vertices, which interpolate between the Witten vertex and the open string lightcone vertex, is also obtained as a flat limit. We use the same limit on the case of closed strings to define the closed string Kaku vertices: a one parameter family of vertices that interpolates between the polyhedral vertices -- which are covariant, but not cubic -- and the closed string lightcone vertex -- which is cubic, but not Lorentz covariant.
\end{abstract}

\vfill
{\renewcommand{\arraystretch}{0.8}
}

\end{titlepage}

\tableofcontents

\numberwithin{equation}{section}

\section{Introduction}

The concept of string field theory (SFT) vertices is a fundamental ingredient in the construction of string field theories \cite{closed_stf_erler}. There are many known choices of string field theory vertices. For open strings, there is the Witten vertex, which consists of a single cubic interaction \cite{witten86}. The lightcone vertex, introduced by Mandelstam \cite{mandelstam73, mandelstam74, lc_mapping}, is another simple choice which consists of a cubic and a quartic interaction. But it is not Lorentz covariant, as the geometry of the interaction depends on the lightcone momenta of the scattered strings. The open string Kaku vertices \cite{kaku88, lc_mapping} are a one parameter family of vertices that interpolate between the open string lightcone vertex and the Witten vertex.

It is known that covariant closed string field theory cannot be cubic \cite{Sonoda:1989sj}. In fact, all known closed SFTs are non-polynomial. A simple covariant choice of vertex is the polyhedral vertices \cite{polyhedra,Kugo:1989aa}. The minimal area vertices \cite{Zwiebach:1990ew,Zwiebach_1993} are closely related to the polyhedral vertices -- they differ only at the quantum level. The closed string lightcone vertex has only a cubic interaction, but is not Lorentz covariant. Other examples of closed string vertices include the $SL(2,\mathbb{C})$ vertices, studied recently in \cite{Erbin:2023rsq}, and the newly investigated vertices for strings on a background with a large number of D-branes \cite{fırat2023string}.

The hyperbolic vertices \cite{costello2019hyperbolic, Cho:2019anu} are a one parameter family of vertices, defined in terms hyperbolic metrics on the Riemann surface. The parameter of the family is a number $L$ related to the length of the scattered strings. The hyperbolic vertices are defined for both closed and open strings. The flat limit of the hyperbolic geometry are the polyhedral vertices, for closed strings, or the Witten vertex, for open strings.

In order to make computations with the vertices, one has to specify what are the associated local coordinate maps of the vertices. For the hyperbolic vertices, the local coordinates of the 3-vertex are computed in \cite{Firat:2021ukc}, where the expressions are generalized for different lengths $L_i$ associated to each scattered string. This motivates the definition of a generalization of the hyperbolic vertices, using different lengths $L_i$. Recently there have been some works in this direction, such as \cite{ishibashi2023fokkerplanck,ishibashi2024strebel}. Other investigation \cite{F_rat_2023} of the hyperbolic vertices states some results on the relation of higher vertices with the cubic vertex in closed SFT.

In this paper, we give a definition of a two parameter family of vertices that generalize the ideas of both the hyperbolic vertices and the Kaku vertices. We refer to this family as the hyperbolic Kaku vertices. It is defined in terms of hyperbolic geometries, as the hyperbolic vertices. But the geometries also depend on state inputs, as is the case in lightcone and Kaku vertices. We give the definition only at tree level, i.e. looking only at the moduli spaces of Riemann surfaces with genus zero.

The definition of hyperbolic Kaku vertices is done for both open and closed strings. In both cases, the two parameter family of vertices contain the hyperbolic vertices as a boundary. For open strings, the flat limit of the hyperbolic Kaku vertices is the open string Kaku vertices. We use the flat limit on the closed string hyperbolic Kaku vertices to obtain the closed string Kaku vertices, and thus characterize a family of vertices which interpolate between the closed lightcone vertices (which are cubic, but not Lorentz covariant), and the polyhedral vertices (which are covariant, but not cubic).

In Section \ref{vertices} we review the relevant choices of vertices for our definition: hyperbolic vertices, lightcone vertices and open string Kaku vertices. In Section \ref{section_hyperbolic_lc_vertices}, we introduce the hyperbolic Kaku vertices and show that they are a consistent definition of vertices. In section \ref{section_flat_limit}, we define precisely what is meant by flat limit. We show that, for open strings, the flat limit gives the open string Kaku vertices. And then we characterize the 4-vertex of the closed string Kaku vertices, interpolating between the lightcone and the polyhedral 4-vertex. We finish with conclusions and comments about future directions in Section \ref{section_conclusion}.

\section{Review of hyperbolic and lightcone vertices}
\label{vertices}

In closed SFT, a choice of string vertex amounts to considering the moduli space $\mathcal{M}_{g,n}$ of Riemann surfaces with $n$ punctures and genus $g$ and specifying which region of it is taken as the interaction vertex of the field theory. Additionally, the vertex should contain information of local coordinate patches on the surfaces. So the $n$-vertex $\mathcal{V}_{g,n}$ is an integration cycle in a section of a bundle over $\mathcal{M}_{g,n}$, where the fibers are choices of local coordinates. The full vertex is given by $\mathcal{V} = \sum_{g,n} \hbar^g \mathcal{V}_{g,n}$, and it satisfies the master equation
\begin{align}
    \partial \mathcal{V} + \hbar \Delta \mathcal{V} + \frac{1}{2} \{ \mathcal{V}, \mathcal{V} \} = 0,
\end{align}
where the operations roughly correspond to the following: $\partial$ is the boundary operator, $\Delta$ removes two local coordinate patches and glues the boundaries, and $\{ \cdot, \cdot \}$ takes two surfaces, removes a local coordinate patch from each, and glue them. These operations are defined in detail in \cite{Sen_1994}.

In this section we review some choices of vertex which are important for our construction of the hyperbolic Kaku vertices: the hyperbolic vertices, the lightcone vertices, and the Kaku vertices.

\subsection{Hyperbolic vertices}

The hyperbolic string vertices \cite{costello2019hyperbolic, Cho:2019anu} are defined in terms of hyperbolic metrics on the Riemann surfaces. First one parameterizes the moduli space of Riemann surfaces with choices of hyperbolic metrics, and then one specifies the $n$-vertex regions in terms of this parameterization. Let's describe briefly how the definition goes for open strings, and then explain how to generalize it for closed strings. We will focus only on the tree level.

\vspace{.1cm}
\noindent{\bf Open strings.} To characterize the $n$-vertex, we look at the moduli space $\mathcal{M}_n^\text{disk}$ of disks with $n$ boundary punctures. Then we can describe how to construct an element of $\mathcal{M}_n^\text{disk}$ with a hyperbolic metric. Many useful properties of hyperbolic geometry are found in \cite{buser1992geometry,beardon1983geometry}.

Consider a surface with the shape of a polygon with $2n$ sides. On it, we consider a hyperbolic metric. For instance, we can take it to be
\begin{align}
    ds^2 = & ~ \frac{dz d \bar z}{(\text{Im}z)^2},
\end{align}
on the UHP. The geodesics are segments of circles centered on the real line, and the metric is invariant under $SL(2,\mathbb{R})$ transformations. We fix the sides of the polygon to be geodesics that meet at right angles. Such a polygon has $2n-3$ degrees of freedom: there are $2n$ dof corresponding to the position of the center of each circle, $2n$ dof corresponding to their radii, and $2n$ constraints by imposing that they meet at right angles. The $SL(2,\mathbb{R})$ invariance of the metric makes $SL(2,\mathbb{R})$-related polygons equivalent, fixing three more degrees of freedom.

To parameterize $\mathcal{M}_n^\text{disk}$, we fix $n$ non-subsequent sides to some fixed length $L_i$, $i=1,\cdots,n$. This leaves the space of hyperbolic polygons with $n-3$ degrees of freedom. Then the surface in $\mathcal{M}_n^\text{disk}$ is defined by attaching infinite flat strips to each side where we fixed the length $L_i$. The flat metric in each strip is chosen in such a way that the metric on the whole surface is continuous. In other words, the boundaries of the strip and the hyperbolic surface are glued in such a way as to match the proper lengths, determined by the flat and hyperbolic metrics respectively. This operation is called {\it grafting}. The infinite flat strip attached to the side of length $L_i$ is the local coordinate patch around the puncture $i$. We call $L_i$ the {\it external lengths}.

\begin{figure}[ht]
    \centering
    \includegraphics[width=\textwidth]{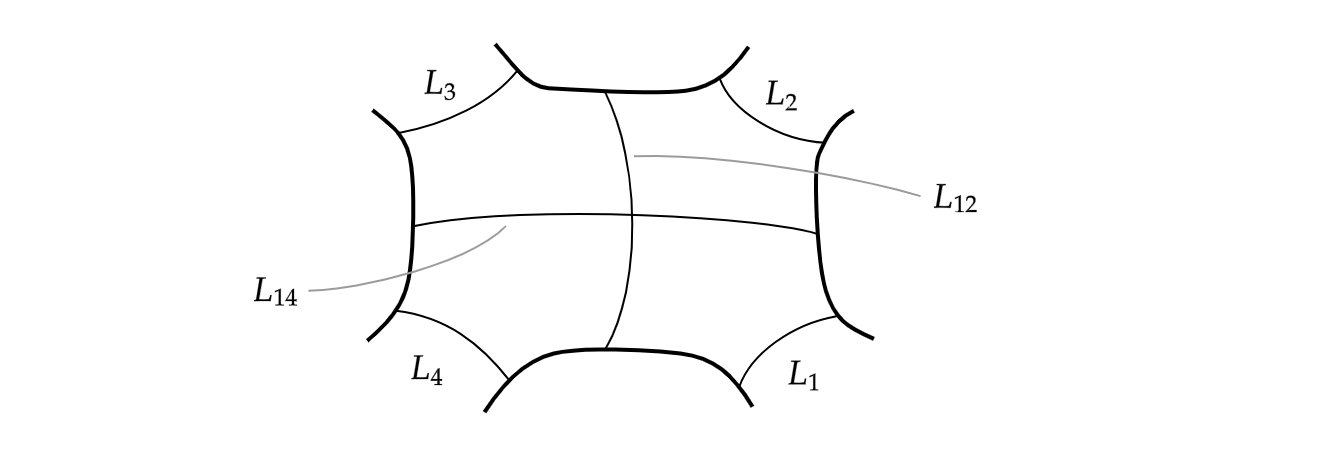}
    \caption{Depiction of the bulk geodesics in the 4-punctured Riemann surface with boundary obtained from the polygon with hyperbolic metric.}
    \label{bulk_geodesics}
\end{figure}

After grafting the local coordinate patches, the sides of the hyperbolic polygon split in two kinds of curves: the ones that are segments along the boundary in between the grafted patches, and the segments of external lengths $L_i$, which are the interface in between the hyperbolic polygon and the local coordinate patches. There are also other geodesics that go through the bulk of the hyperbolic polygon and which are orthogonal to the boundaries of the surface. We call them {\it bulk geodesics}. Each bulk geodesic splits the punctures in two groups. So it defines a scattering channel. Consider a bulk geodesic that separates the (subsequent) punctures $1, 2, \cdots, j$ from the punctures $j+1, \cdots, n-1, n$. We denote this geodesic\footnote{By abuse of notation, we use the same symbol to denote both the geodesic and its length.} as $L_{1\cdots j}$. Figure \ref{bulk_geodesics} shows the bulk geodesics for $n=4$. There are four external lengths (they are $L_1$, $L_2$, $L_3$ and $L_4$), and two bulk geodesics, $L_{12}$ and $L_{14}$, representing the $s$- and $t$-channels.

For the definition of the hyperbolic vertices, we consider the parameterization of $\mathcal{M}_n^\text{disk}$ where all external lengths $L_i$ are equal to the same length $L$. The condition for a point of $\mathcal{M}_n^\text{disk}$ to be inside the $n$-vertex region is a bound on the systole. The systole of a surface is the bulk geodesic of shortest length. Let's say that $S$ is the length of the systole. Then the $n$-vertex condition is
\begin{align}
    S \geq L.
\end{align}
In other words, a surface where all the non-contractible geodesics are bigger than $L$ is a part of the $n$-vertex. We can substitute the $n$-vertex condition to a bound on every bulk geodesic:
\begin{align}
    L_\mathcal{I} \geq L,
\end{align}
for every $\mathcal{I}$ which is a list of subsequent punctures. When $\mathcal{I} = i$ consists of a single puncture, the bound saturates to an equality, that defines the external length $L_i$.

\vspace{.1cm}
\noindent{\bf Closed strings.}
To define the hyperbolic vertices for closed strings, we need to parameterize the moduli space $\mathcal{M}_n$ of spheres with $n$ punctures. We will characterize a surface in $\mathcal{M}_n$ as a pair of hyperbolic polygons with $2n$ sides each. The space of hyperbolic metrics on each polygon is $2n-3$, so the total number of degrees of freedom is $4n-6$. We then glue the two polygons by non-subsequent sides. But the glued sides must have the same lengths, so we impose $n$ gluing consistency constraints. We obtain a surface with $n$ holes, and we fix their perimeters to the fixed external lengths $L_i$, $i=1,\cdots,n$. We are left with $2n-6$ degrees of freedom, which is the dimensionality of $\mathcal{M}_n$. Then the process of grafting consists in attaching an infinite flat cylinder to each hole. This defines the local coordinate patches.

For closed strings, we use the term bulk geodesic to refer to a closed geodesic on the bulk (away from the local coordinate patches). There are infinitely many homotopically different closed curves that define the same channel, i.e. that split the punctures into the same two groups. For each homotopy class, there is a unique geodesic. So there are infinitely many closed geodesics that define the same channel.

The hyperbolic vertices for closed strings are again defined by fixing the external length $L_i$ around all local coordinate patches to be the same, $L_i = L$. Then the $n$-vertex condition is a bound on the systole $S$ (the shortest non-contractible geodesic), given by $S \geq L$.

\subsection{Lightcone vertices}

The lightcone vertex \cite{mandelstam73,mandelstam74,lc_mapping} depends on inputs of the states. The $n$-vertex is a function of $n$ real numbers $\alpha_i$ which satisfy a conservation equation
\begin{align}
    \sum\limits_{i=1}^n \alpha_i = 0.
\end{align}
The inputs can have different signs, so they can be split between incoming and outgoing states. The lightcone vertex appears in the context of lightcone SFT, where the inputs are the string lengths, which are proportional to the component $k^-_i$ of the momenta \cite{lc_mapping}.

\vspace{.1cm}
\noindent{\bf Open string lightcone vertices.} To define the lightcone vertices, we parameterize the $n$-punctured open string Riemann surface as a Mandelstam diagram. Suppose that the $n$ inputs $\alpha_i$ are split in $j$ incoming and $n-j$ outgoing states. Then the Mandelstam diagram is an infinite flat surface consisting of $j$ parallel strips of lengths $\alpha_i$, $i=1,\cdots,j$, that join and split at different points, and turn into $n-j$ parallel strips of lengths $\alpha_i$, $i=j+1,\cdots,n$.

\begin{figure}[ht]
    \centering
    \includegraphics[width=\textwidth]{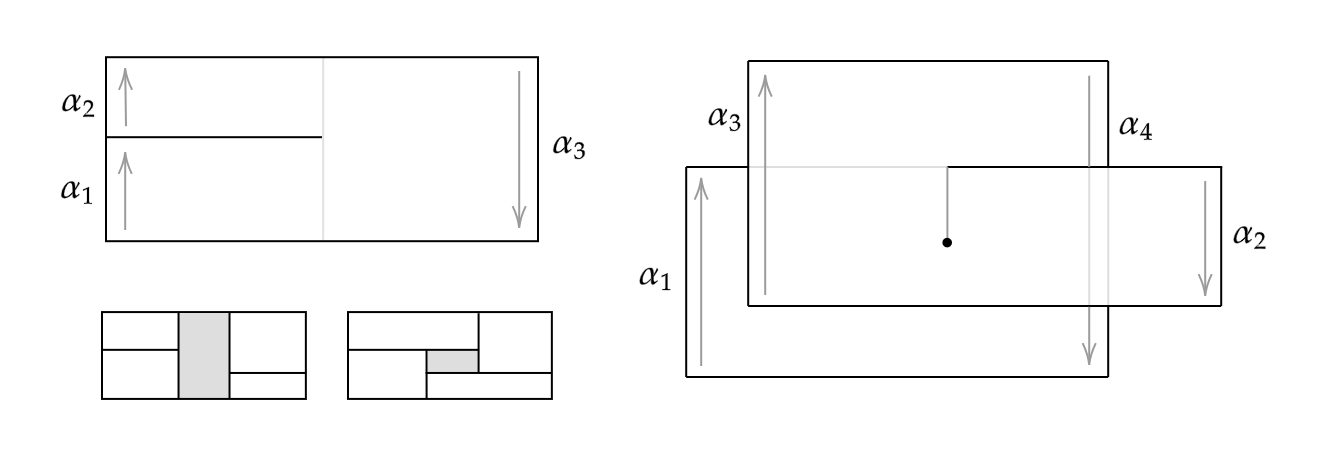}
    \caption{On the left and above: the Mandelstam diagram corresponding to the open string lightcone 3-vertex. Below it, examples of $s$- and $t$-channel diagrams produced by attaching two 3-vertices with propagators. On the right: the non-planar Mandelstam diagram corresponding to the lightcone 4-vertex for open strings. The point in the bulk of the diagram is the position of the conical singularity.}
    \label{mandelstam_diagrams}
\end{figure}

The 3-vertex is the Mandelstam diagram with three strips (one incoming and two outgoing, or equivalently two incoming and one outgoing). Almost every Mandelstam diagram with four strips can be obtained by joining two 3-vertices through a propagator, unless the inputs are in the so-called {\it non-planar configuration}. If the punctures associated with inputs $\alpha_1$, $\alpha_2$, $\alpha_3$ and $\alpha_4$ are displaced in this order along the boundary, but their signs in the conservation equation are alternating, e.g. if it is of the form $|\alpha_1| + |\alpha_3| = |\alpha_2| + |\alpha_4|$, then there are some Mandelstam diagrams that cannot be obtained by joining 3-vertices with propagators. Thus we also have a 4-vertex, which has non-planar geometry. There is a one parameter family of non-planar diagrams defining the 4-vertex, and they can be parameterized by the position of a conical singularity in the bulk. The surface has a flat metric everywhere, except in the point of conical singularity. The boundary of the 4-vertex corresponds to the surfaces where the conical singularity hits the boundary of the diagram. Figure \ref{mandelstam_diagrams} shows the geometry of the 3-vertex and the 4-vertex, as well as the structure of $s$- and $t$-channels in the planar geometries.

\vspace{.1cm}
\noindent{\bf Closed string lightcone vertex.}
The closed string Mandelstam diagram is defined as the gluing of two open string Mandelstam diagrams. Because there is no issue of puncture ordering, there is no necessity of a 4-vertex for any momentum configuration, so the lightcone vertex for closed strings is cubic.

\subsection{Kaku vertices}

\begin{figure}[ht]
    \centering
    \includegraphics[width=.8\textwidth]{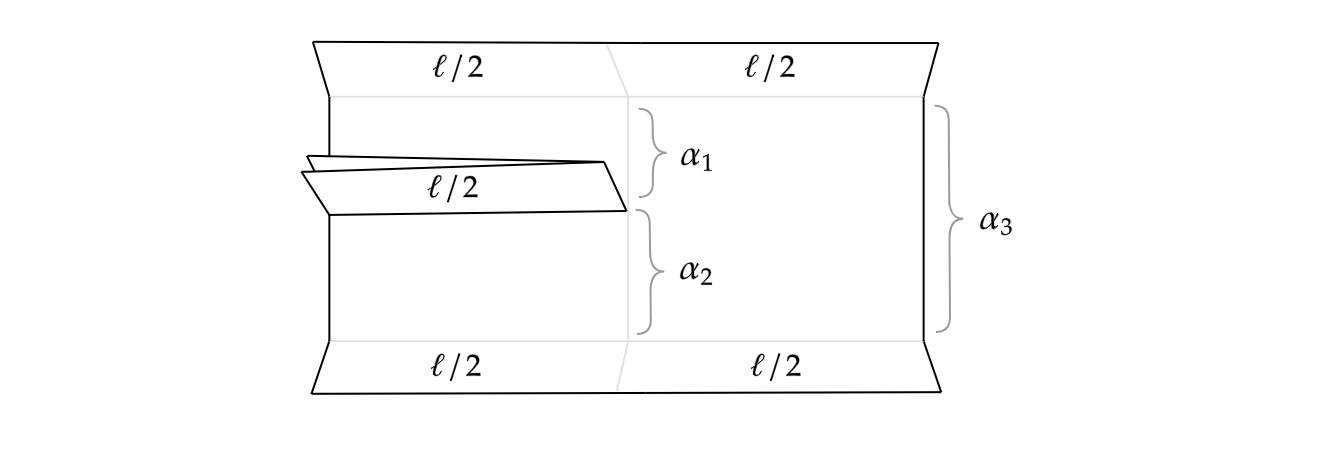}
    \caption{Geometry of the Kaku 3-vertex.}
    \label{kaku-vertex}
\end{figure}

\noindent{\bf Open string Kaku vertices.}
There is a family of vertices that interpolate between the open string lightcone vertices and the Witten vertex. It is known as Kaku vertices \cite{kaku88,lc_mapping}. They are obtained by a deformation of the open string lightcone vertices. The 3-vertex geometry is defined by attaching Chan-Paton strips of length $\ell/2$ at both sides of each strip of the Mandelstam diagram that defines the lightcone 3-vertex. Figure \ref{kaku-vertex} shows the geometry of the 3-vertex. The 4-vertex is also defined by taking the lightcone 4-vertex and attaching Chan-Paton strips of length $\ell/2$ to each side of each strip.

In the limit $\ell \to \infty$, the strips of the 3-vertex approach the same length, so we obtain the Witten 3-vertex upon a rescaling of the metric by $1/\ell$. The non-planar 4-vertex consists of all surfaces obtained by varying the position of the conical singularity from one Chan-Paton strip to the other, but it does not go over to the Chan-Paton strip. As $\ell\to \infty$, the interval covered by the conical singularity shrinks to a point, so the 4-vertex vanishes. Thus the $\ell\to\infty$ limit of the Kaku vertex is the Witten vertex.

\vspace{.1cm}
\noindent{\bf Closed string Kaku vertices.} As far as the authors know, there is no definition of the closed string Kaku vertices. In analogy to the open string Kaku vertices, they should be an interpolation between the closed lightcone vertex and the polyhedral vertices.

We note that an open-closed Kaku SFT has been defined in \cite{ando2023closed}.\footnote{We thank Nathan Berkovits for bringing this work to our attention.} However, the theory has open string interactions given by open string Kaku vertices, while the closed string interactions are given by closed string lightcone vertices.

We want to define the closed string Kaku vertices in the following way. Consider a deformation of the closed string Mandelstam diagram by changing the circumference of the incoming cylinders from $\alpha_i$ to $\alpha_i + \ell$, where $\alpha_i$ are the state inputs. The change of the circumference by $\ell$ is the closed string analogous of the addition of Chan-Paton strips in the open string diagrams. We expect that in the $\ell \to \infty$ limit we get the polyhedral vertices, which have non-vanishing $n$-vertex contributions for all $n \geq 3$. If that is the case, then the closed string Kaku vertex for some finite $\ell$ should contain non-vanishing $n$-vertices for all $n\geq 3$ as well.

This seems like an obstruction if one wants to define the closed string Kaku vertices as a deformation of the lightcone vertices. One possible approach to understand this interpolation, as we will see, is to have the vertices defined in terms of critical graphs, similarly to the polyhedral vertices.

In the next section, we will instead define a two parameter space of vertices which has features of both the hyperbolic vertices and the lightcone vertices. The definition will be made for both open and closed strings. For open strings, we will show that the space contain the open string Kaku vertex as a limit. Then we will take the analogous limit in the space of vertices defined for closed strings, and we will find the closed string Kaku vertices defined in terms of critical graphs.

\section{Hyperbolic Kaku vertices}
\label{section_hyperbolic_lc_vertices}

Here we give a definition of a family of string field theory vertices which is similar to both the hyperbolic vertices and the lightcone vertices. The definition is applied to both open and closed strings. The vertex depends on two real parameters, which we will take to be $(r,\theta)$. These parameters define polar coordinates in the two dimensional space of vertices. The domain is
\begin{align}
    r \geq 0, \hspace{.6cm} 0 \leq \theta \leq \pi/2.
\end{align}
The vertex at the point $(r,\theta)$ is defined through the parameterization of the moduli spaces $\mathcal{M}_n$ in terms of hyperbolic metrics. The geometry of the vertex depends on state inputs $\alpha_i$. Let's give the definition first for open strings and then to closed strings.

\vspace{.1cm}
\noindent{\bf Open strings.}
Consider a point in $\mathcal{M}_n^\text{disk}$ defined in terms of a hyperbolic metric in a $2n$-sided polygon, with grafted local coordinate patches on $n$ non-subsequent sides. In the same way that the parameterization was described when introducing the hyperbolic vertices, we take the geodesics around each local coordinate patch to have fixed external lengths $L_i$. But differently from the hyperbolic vertices, where in the end they would all be fixed to the same length $L$, in this family of vertices we fix the external lengths to be
\begin{align}
    L_i = & ~ r ( \cos \theta + | \alpha_i | \sin \theta).
\end{align}
The $n$-vertex region is defined using bounds on every geodesic. For a channel that splits the collection of punctures $\mathcal{I} = i_1 \cdots i_j $ from the other ones, we define the incoming momentum as
\begin{align}
    \alpha_\mathcal{I} = & ~ \sum_{m = 1}^j \alpha_{i_{m}}.
\end{align}
Note that the momenta $\alpha_{i_{m}}$ might carry different signs.  The condition for the surface to be part of the $n$-vertex are the bounds
\begin{align}
    L_\mathcal{I} \geq r ( \cos \theta + | \alpha_\mathcal{I} | \sin \theta).
\end{align}
If the bounds are satisfied for every channel $\mathcal{I}$, the surface is in the $n$-vertex.

At $\theta = 0$ we have the hyperbolic vertices, with the identification $L=r$. The geodesics bounds are given by
\begin{align}
    L_\mathcal{I} \geq r \equiv L,
\end{align}
which is exactly the definition of the hyperbolic vertices.

At $\theta = \pi/2$, the geodesic bounds are given by
\begin{align}
    L_\mathcal{I} \geq r | \alpha_\mathcal{I} |.
\end{align}
The hyperbolic geometry of these vertices resembles the geometry of the Mandelstam diagrams, in the sense that the external lengths satisfy a conservation equation
\begin{align}
    \sum_{i=1}^n L_i = & ~ \sum_{i=1}^n r \alpha_i = 0
\end{align}
at each level $n\geq 3$. So this family of vertices is a hyperbolic generalization of the lightcone vertices. The $r\to \infty$ limit of this family is the open string lightcone vertex, as we will prove later.

\begin{figure}[ht]
    \centering
    \includegraphics[width=.9\textwidth]{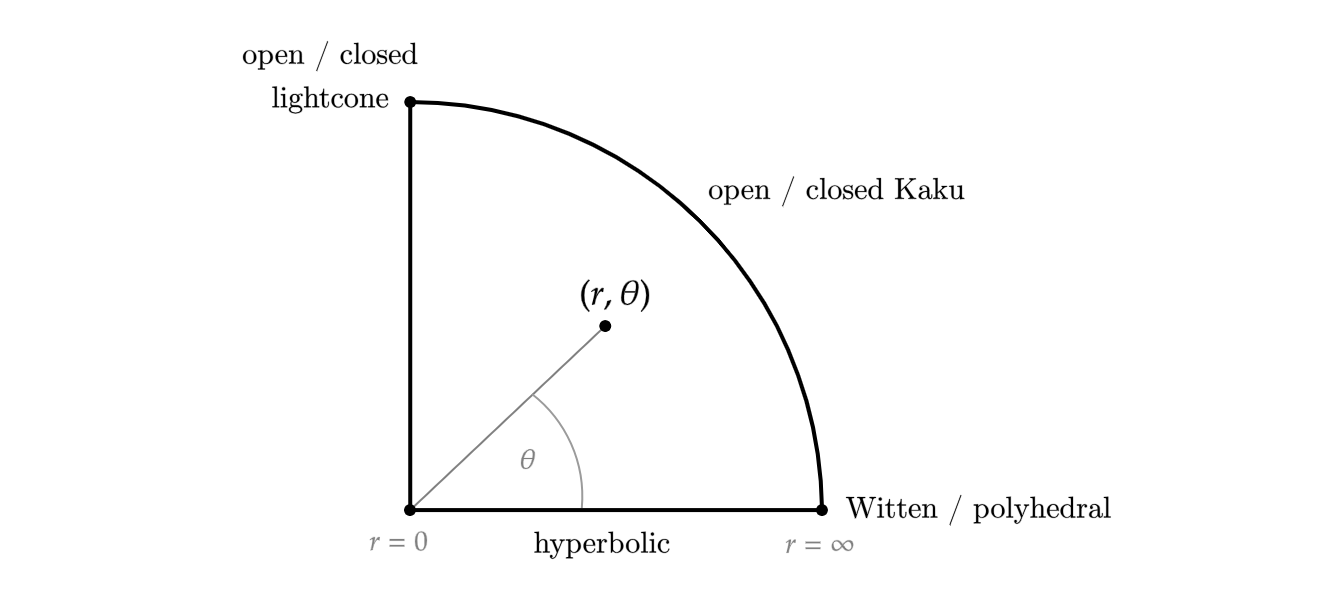}
    \caption{The space of hyperbolic Kaku vertices, parameterized by $(r,\theta)$.}
    \label{hyperbolic-kaku-vertices}
\end{figure}

\vspace{.1cm}
\noindent{\bf Closed strings.} The generalization of the definition for closed strings is completely analogous to the generalization made in the case of the hyperbolic vertices. One defines the hyperbolic metric in two $2n$-sided polygons, glue them together, and attach flat cylinders to define the local coordinates. There are infinitely many geodesics for each channel which splits more than one puncture. The expression for the external lengths remains the same, $L_i = r(\cos \theta + \alpha_i \sin \theta)$. For each channel $\mathcal{I}$, it is enough to impose the bound on the shortest geodesic $L_\mathcal{I}$. The bounds are $L_\mathcal{I} \geq r (\cos \theta + \alpha_i \sin \theta)$.

We will refer to this two parameter space of vertices as {\it hyperbolic Kaku vertices}. Figure \ref{hyperbolic-kaku-vertices} shows the space of vertices for both open and closed strings. We claim that, for open strings, the boundary at $r\to \infty$ corresponds to the open string Kaku vertices. This family of vertices interpolates between the Witten vertex, at $(r,\theta) = (\infty, 0)$, and the open string lightcone vertex, at $(r,\theta) = (\infty, \pi/2)$. For closed strings, we will take the $r\to \infty$ limit as a definition of the closed string Kaku vertices. This will be the family of vertices that interpolates between the polyhedral vertices, at $(r,\theta) = (\infty, 0)$, and the close string lightcone vertex, at $(r,\theta) = (\infty, \pi/2)$.

The local coordinate maps for the 3-vertex are in \cite{Firat:2021ukc}, where the local coordinate maps for the hyperbolic vertex are computed and the expressions are generalized for arbitrary external lengths $L_i$.

\subsection{Consistency of the geodesic bounds}

The geodesic bounds give a consistent definition of the vertices. To show that, we must prove the master equation in two directions. We will prove it for the case of open strings. For closed strings, the proof is analogous.

Consider first a surface at the boundary of the $n$-vertex. Then one of the geodesic bounds is satisfied as an equality. We say that the bound is saturated. By cutting the surface through the given geodesic, we generate two surfaces: one with $n-k$ punctures, and another with $k+2$ punctures, for some $1 \leq k \leq n -3$. The surfaces are contained in the $(n-k)$-vertex and in the $(k+2)$-vertex, respectively.

\begin{figure}[ht]
    \centering
    \includegraphics[width=.9\textwidth]{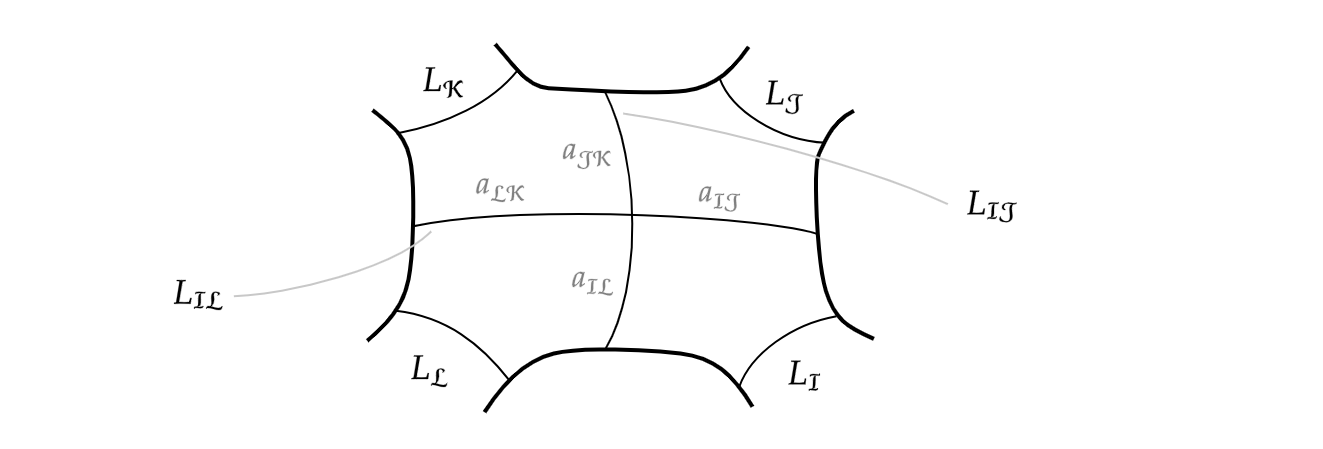}
    \caption{Generalization of Figure \ref{bulk_geodesics}, where the external lengths are exchanged for arbitrary geodesics, which can represent either an external state or a scattering channel.}
    \label{bound_consistency}
\end{figure}

For the other direction, we want to show that upon joining an $n$-vertex and an $m$-vertex surfaces we obtain a surface contained in the $(n+m-2)$-vertex -- that is, all the bounds are satisfied in the resulting surface. For all the geodesics that lie within one of the initial surfaces, the corresponding bounds are unmodified, and since the initial surfaces are vertex surfaces, these bounds are satisfied.

Now let's consider the geodesics that cross between the two surfaces. In Figure \ref{bound_consistency}, we call $L_{\mathcal{I}  \mathcal{J} }$ the side by which the surfaces are joined, and we want to analyse the bound on the geodesic $L_{\mathcal{I}  \mathcal{L}}$. The two geodesics divide the external legs into four sets, which we label by $\mathcal{I,J,K,L}$. The geodesics $L_\mathcal{I}, L_\mathcal{J}, L_\mathcal{K}, L_\mathcal{L}$ isolate each of these sets.

The geodesics $L_{\mathcal{I}  \mathcal{J}}$ and $L_{\mathcal{I}  \mathcal{L}}$ cross each other once. We split the curves through the points where they touch, and define four curves with lengths $a_{\mathcal{I}  \mathcal{J}}$, $a_{\mathcal{J}  \mathcal{K}}$, $a_{\mathcal{K}  \mathcal{L}}$ and $a_{\mathcal{I}  \mathcal{L}}$. Now consider the curve obtained by joining $a_{\mathcal{I}  \mathcal{J}}$ and $a_{\mathcal{J}  \mathcal{K}}$. This is a curve homotopic to $L_\mathcal{J}$. By virtue of the hyperbolic metric, a curve homotopic to a geodesic cannot be shorter than said geodesic. Thus, $a_{\mathcal{I}  \mathcal{J}}+a_{\mathcal{J}  \mathcal{K}}\geq L_\mathcal{J}$, and similarly for other pairs of $a$ curves. This implies
\begin{align}
    L_{\mathcal{I}  \mathcal{L}} + L_{\mathcal{I}  \mathcal{J}} \geq & ~ L_\mathcal{J} + L_\mathcal{L}, \nonumber \\
    L_{\mathcal{I}  \mathcal{L}} + L_{\mathcal{I}  \mathcal{J}} \geq & ~ L_\mathcal{I} + L_\mathcal{K}.
\end{align}
Since $L_{\mathcal{I}  \mathcal{J}}$ is the joining curve, $L_{\mathcal{I}  \mathcal{J}} = |\alpha_{\mathcal{I}  \mathcal{J}}| + \ell$. Also, since $L_\mathcal{I}, L_\mathcal{J}, L_\mathcal{K}, L_\mathcal{L}$ are each contained within one of the initial surfaces, they satisfy the corresponding bounds. Thus, we get
\begin{align}\label{bounds}
    L_{\mathcal{I}  \mathcal{L}} \geq & ~ |\alpha_\mathcal{J}| + |\alpha_\mathcal{L}| - |\alpha_{\mathcal{I}  \mathcal{J}}| + \ell, \nonumber \\
    L_{\mathcal{I}  \mathcal{L}} \geq & ~ |\alpha_\mathcal{I}| + |\alpha_\mathcal{K}| - |\alpha_{\mathcal{I}  \mathcal{J}}| + \ell.
\end{align}

We want to show that $L_{\mathcal{I}  \mathcal{L}} \geq |\alpha_{\mathcal{I}  \mathcal{L}}|+\ell$. This will clearly follow from \ref{bounds} if either of the following inequalities on the momenta hold:
\begin{align}\label{bounds momenta}
    |\alpha_{\mathcal{I}  \mathcal{L}}| \leq & ~ |\alpha_\mathcal{J}| + |\alpha_\mathcal{L}| - |\alpha_{\mathcal{I}  \mathcal{J}}|, \quad \text{or} \nonumber \\
    |\alpha_{\mathcal{I}  \mathcal{L}}| \leq & ~ |\alpha_\mathcal{I}| + |\alpha_\mathcal{K}| - |\alpha_{\mathcal{I}  \mathcal{J}}|.
\end{align}
To show this, we consider two possible cases, depending on the signs of the momenta. Without loss of generality, assume $|\alpha_\mathcal{I}|\geq|\alpha_\mathcal{J}|,|\alpha_\mathcal{K}|,|\alpha_\mathcal{L}|$. In this case, we will show that the second inequality is always satisfied. We have
\begin{itemize}
    \item If $\alpha_\mathcal{I}$ and $\alpha_\mathcal{L}$ have the same sign, that implies that $|\alpha_\mathcal{K}|$ and $|\alpha_\mathcal{J}|$ have the opposite sign, i.e. $|\alpha_\mathcal{I}|-|\alpha_\mathcal{J}|+|\alpha_\mathcal{L}|-|\alpha_\mathcal{K}|=0$. Then  $|\alpha_{\mathcal{I}  \mathcal{L}}| = |\alpha_\mathcal{I}|+|\alpha_\mathcal{L}| = |\alpha_\mathcal{J}|+|\alpha_\mathcal{K}|$. On the other hand,  $|\alpha_{\mathcal{I}  \mathcal{J}}| = |\alpha_\mathcal{I}|-|\alpha_\mathcal{J}|$. Adding $0=|\alpha_\mathcal{J}|+|\alpha_\mathcal{K}|-|\alpha_{\mathcal{I}  \mathcal{L}}|$ on the right hand side, we get 
    \begin{equation}
        |\alpha_{\mathcal{I}  \mathcal{J}}| = |\alpha_\mathcal{I}|+|\alpha_\mathcal{K}|-|\alpha_\mathcal{IL}|,
    \end{equation}
    which is just the saturated case of the second inequality in \ref{bounds momenta}.
    \item If $\alpha_\mathcal{I}$ and $\alpha_\mathcal{L}$ have opposite signs, then $|\alpha_{\mathcal{I}  \mathcal{L}}| = |\alpha_\mathcal{I}|-|\alpha_\mathcal{L}|$. Now we use the triangular inequality $|\alpha_{\mathcal{I}  \mathcal{J}}| = |\alpha_{\mathcal{K}  \mathcal{L}}| \leq |\alpha_\mathcal{K}|+|\alpha_\mathcal{L}|$. Adding $0 = |\alpha_\mathcal{I}|-|\alpha_\mathcal{L}|-|\alpha_{\mathcal{I}  \mathcal{L}}|$ on the right hand side, we get
    \begin{equation}
        |\alpha_{\mathcal{I}  \mathcal{J}}| \leq |\alpha_\mathcal{K}|+|\alpha_\mathcal{I}|-|\alpha_\mathcal{IL}|,
    \end{equation}
    which again recovers \ref{bounds momenta}.
\end{itemize}
So if $|\alpha_\mathcal{I}|$ is the largest input, the second inequality in \ref{bounds momenta} is always satisfied. This proves that upon gluing two vertex surfaces, the resulting surface is also in a vertex. The master equation is thus satisfied.

For the closed string, there are infinitely many closed geodesics for each channel (one for each homotopy class). Any pair of closed geodesics will cross at least twice, and divide the external states into four sets. It is always possible to form, from pieces of the two closed geodesics, four closed curves isolating each of the four sets. Each of these curves is necessarily longer than the closed geodesic to which it is homotopic, and which satisfies the appropriate bound. From here, the proof is identical to the one for the open string.

\section{The flat limit}
\label{section_flat_limit}

Let's characterize the vertices in the $r\to \infty$ limit. In this limit, all the sides $L_i$ go to infinity. So to analyse the geometry of the Riemann surface in this limit it is interesting to first rescale the metric by $1/r$ (or by any factor inversely proportional to $r$) and then take $r\to \infty$. Before the rescaling, the metric in the $2n$-sided polygon has curvature $-1$. After the rescaling, the curvature is $-1/r$. And as $r\to \infty$, the curvature vanishes. So we call this the {\it flat limit}.

\begin{figure}[ht]
    \centering
    \includegraphics[width=.9\textwidth]{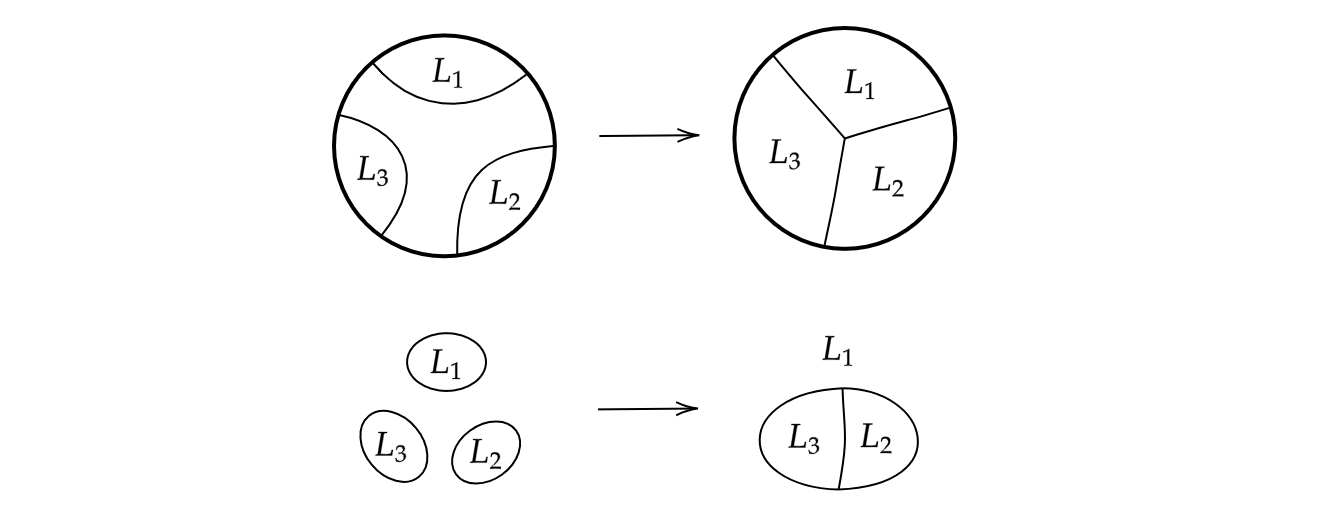}
    \caption{Visualization of the flat limit for $n=3$ for open strings (above) and closed strings (below).}
    \label{polyhedral_limit}
\end{figure}

To make contact with the definition of the Kaku vertices, we define the Chan-Paton strip width $\ell$ as
\begin{align}
    \ell = \cot \theta.
\end{align}
And then we consider a rescaling by $(r \sin \theta)^{-1}$.
In this limit, the shape of the hyperbolic polygons is deformed, and their area goes to zero. The polygons become the union of 1-dimensional manifolds on the Riemann surface. We call this 1-dimensional structure as a {\it critical graph} on the Riemann surface. In the case of closed strings, each critical graph draws a polyhedron on the surface. This makes contact with the language used to define polyhedral vertices. Figure \ref{polyhedral_limit} gives a depiction of how the hyperbolic polygons are deformed into critical graphs on the Riemann surface, at $n=3$, for both open and closed strings.

The external lengths become the perimeter of the faces, and they are given by
\begin{align}
    L_i = | \alpha_i | + \ell.
\end{align}
A geodesic defining a channel $\mathcal{I}$ becomes a set of edges of the critical graph. These edges jointly form a curve that isolates the faces associated with the channel. We refer to the length of this curve as $L_\mathcal{I}$. The geodesic bounds become
\begin{align}
    L_\mathcal{I} \geq | \alpha_\mathcal{I} | + \ell,
\end{align}
which are bounds on the channel curves.

\begin{figure}[ht]
    \centering
    \includegraphics[width=.8\textwidth]{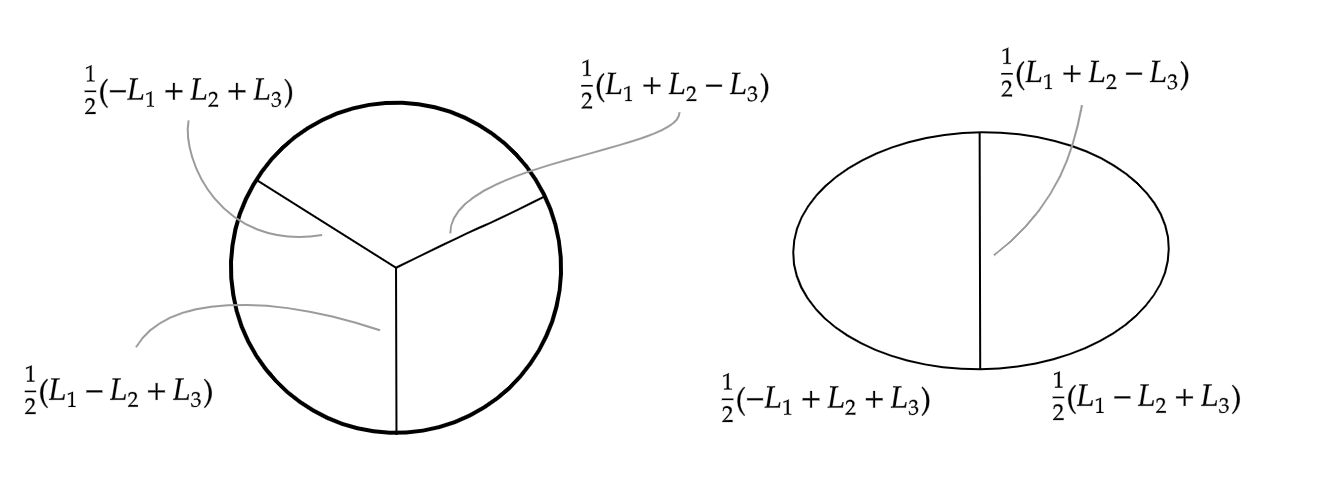}
    \caption{On the left, the critical graph for open strings at $n=3$. On the right, the critical graph for closed strings at $n=3$.}
    \label{3-triangulations-patches}
\end{figure}

Strebel's theorem \cite{polyhedra,Kugo:1989aa} states that the moduli space $\mathcal{M}_n$ of Riemann surfaces with $n$ punctures is isomorphic to the moduli space of critical graphs on the surface. At $n=3$, the moduli space $\mathcal{M}_3$ consists of a single diagram. The theorem also holds for $\mathcal{M}_n^\text{disk}$. Figure \ref{3-triangulations-patches} shows the diagrams that represent $\mathcal{M}_3^\text{disk}$ and $\mathcal{M}_3$ for the choice of parameterization with external lengths $L_1$, $L_2$ and $L_3$. Both diagrams arise from the flat limit of the hyperbolic geometries with the same puncture lenghts $L_1$, $L_2$ and $L_3$. See Appendix \ref{appendix-critical-graphs} for a detailed description of the critical graphs at $n=4$ for open and closed strings.

\begin{figure}
    \centering
    \includegraphics[width=.9\textwidth]{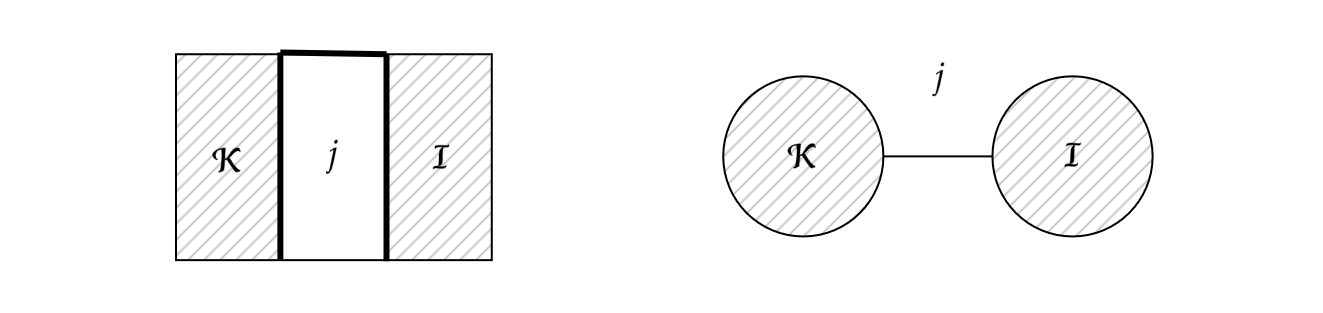}
    \caption{Depiction of a reducible graph of open strings (left) and of closed strings (right).}
    \label{reducible-graphs}
\end{figure}

A critical graph is said to be {\it reducible} if by removing one of the faces, we obtain two critical graphs connected by a single edge. The critical graphs which are not reducible are called {\it irreducible}. For open strings, the irreducible graphs are the ones where all consecutive faces meet. For closed strings, the irreducible graphs are 1PI diagrams. In the following sections, we will show that, in the flat limit, the Kaku vertices are contained in the irreducible graphs. The depictions of a reducible graph for open and closed strings are shown in Figure \ref{reducible-graphs}. The graphs are decomposed into critical graphs corresponding to channels $\mathcal{I}$ and $\mathcal{K}$ (shown as a shaded face), connected with a face $j$.

The {\it polyhedral vertices} \cite{polyhedra,Kugo:1989aa} are defined in terms critical graphs on closed string Riemann surfaces. The external lengths are fixed as $L_i = 2\pi$, $i=1,\cdots, n$, where $n$ is the number of faces of the polyhedron. The $n$-vertex region is defined in terms of bounds on every closed curve:
\begin{align}
    a_p \geq 2\pi,
\end{align}
for $p$ labeling the different closed curves of the polyhedron. The analog of the polyhedral vertices for open strings is the Witten vertex.

We will use the parameterization of the moduli space defined by the critical graphs to analyse the vertices in the flat limit. We will separate the analysis for open and closed strings. For open strings, we will prove that vertices in the flat limit are the open string Kaku vertices. Then we will take the flat limit for closed strings, and we will find the closed string Kaku vertices defined in terms of critical graphs.

\subsection{Open string Kaku vertices}
\label{section_open_string}

Here we derive the open string Kaku vertices as the flat limit of the hyperbolic Kaku vertices. The bounds of the open string Kaku vertices are
\begin{align}\label{geodesic bounds}
    L_\mathcal{I} \geq | \alpha_i | + \ell.
\end{align}

First, let's show that the reducible graphs never contribute to the vertices -- that is, they cannot satisfy the bounds. Consider the open reducible graph in Figure \ref{reducible-graphs}. Note that the two edges on either side of the $j$-th face isolate the $\mathcal{I}$ and $\mathcal{K}$ regions. As such, the graph can only contribute to the vertex if these edges satisfy the corresponding bounds
\begin{align}\label{bounds reducible}
    L_\mathcal{I} \geq |\alpha_\mathcal{I}| + \ell, \hspace{.6cm}
    L_\mathcal{K} \geq |\alpha_\mathcal{K}| + \ell.
\end{align}
But note that the perimeter $L_j$ of the $j$-th face is longer than the sum of these two edges. This implies
\begin{align}
    L_\mathcal{I}+L_\mathcal{K} \leq L_j = |\alpha_j| + \ell = |\alpha_\mathcal{I}+\alpha_\mathcal{K}| + \ell \leq |\alpha_\mathcal{I}| + |\alpha_\mathcal{K}| + \ell
\end{align}
This is only consistent with the bounds \ref{bounds reducible} if $\ell=0$, and if the edge along the boundary of the graph collapses. Therefore, the reducible graphs don't contribute to the vertices.

Now we look at the irreducible graphs. We'll show that the Kaku vertices are at most quartic for all $\ell$ (and cubic in the $\ell\rightarrow\infty$ limit). To do this, it is enough to analyse the bounds for $\mathcal{I}= i j$ and $\mathcal{I}= i j k$ -- i.e. the geodesics isolating either two or three faces.

\begin{figure}[ht]
    \centering
    \includegraphics[width=.9\textwidth]{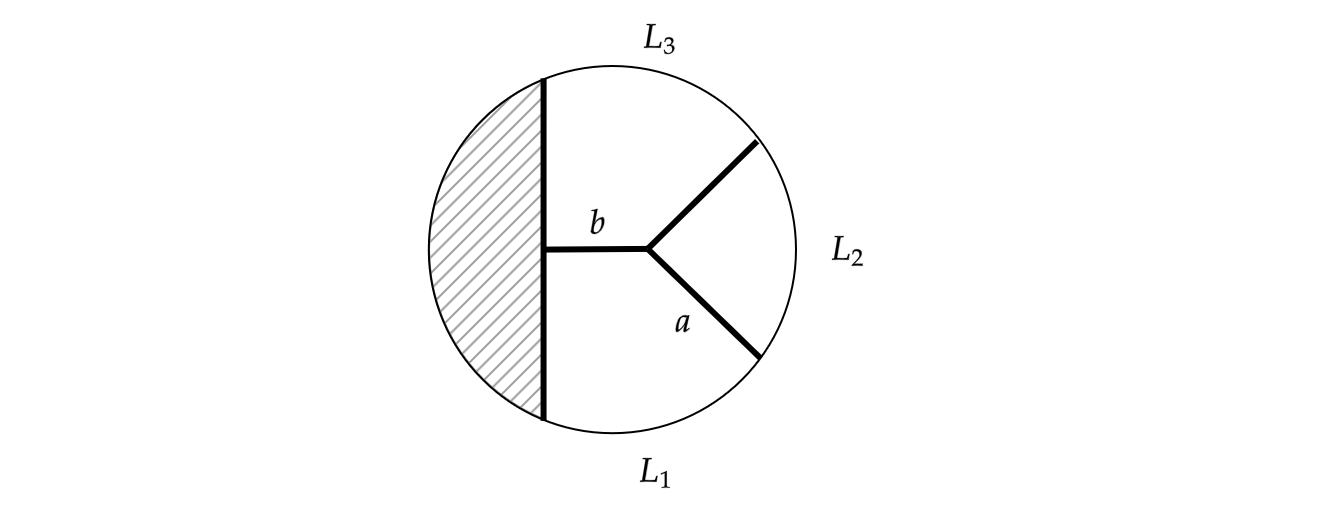}
    \caption{Arbitrary critical graph for open strings, with punctures with lengths $L_1$, $L_2$ and $L_3$.}
    \label{no-planar-vertex}
\end{figure}

First, we choose three consecutive faces $L_1$, $L_2$, $L_3$. Note that, for any irreducible graph, it's always possible to choose a face $L_2$ which only shares edges with $L_1$ and $L_3$, as in Figure \ref{no-planar-vertex}. Now consider the bounds on $L_{12}$ and $L_{23}$:
\begin{align}
    L_{12} = & ~ | \alpha_2 | + | \alpha_1 | + 2\ell - 2a \geq | \alpha_2 + \alpha_1 | + \ell, \nonumber \\
    L_{23} = & ~ | \alpha_3 | - | \alpha_2 | + 2a \geq | \alpha_3 + \alpha_2 | + \ell.  
\end{align}
Now suppose $\alpha_1$ and $\alpha_2$ have the same sign. Then the first bound becomes $a \leq \ell/2$, while the second bound implies $a \geq \ell/2$, regardless of the signs. Then, the parameter $a$ is fixed at $a=\ell/2$, which means the vertex region degenerates. Thus, a vertex can only appear if the signs are alternating on the three faces.

Next, consider the geodesic $L_{123}$ isolating these three faces, if it exists -- that is, if it's a five-vertex or higher. The bound is 
\begin{align}
    L_{123} = | \alpha_1 | + | \alpha_3 | - | \alpha_2 | + \ell - 2b \geq | \alpha_3 + \alpha_2 + \alpha_1 | + \ell. 
\end{align}
If the signs are alternating, then $| \alpha_1 | + | \alpha_3 | - | \alpha_2 | = \pm | \alpha_3 + \alpha_2 + \alpha_1 |$. The bound only has a solution (with non-negative $b$) if $| \alpha_1 | + | \alpha_3 | - | \alpha_2 | = | \alpha_3+\alpha_2+\alpha_1 |$, in which case it implies $b = 0$. Thus, the vertex is degenerate even when the signs alternate. This shows that there are no vertices higher than four. A 4-vertex can exist -- since in this case the bound on $L_{123}$ does not apply -- but only if the signs are alternating. This is in accordance with the Kaku and lightcone vertices as defined in \cite{lc_mapping}. The tetrahedron with alternating signs corresponds to the non-planar Mandelstam diagrams.

Let us analyse the diagram with alternating signs and see that there is indeed a 4-vertex. The geodesic bounds are
\begin{align}
    L_{12} = & ~  \alpha_1 + | \alpha_2 | + 2\ell - 2a \geq \alpha_1 - | \alpha_2 | + \ell \implies a \leq | \alpha_2 | + \ell/2, \nonumber \\
    L_{14} = & ~ \alpha_3 - | \alpha_2 | + 2a \geq \alpha_3 - | \alpha_2 | + \ell \implies a \geq \ell/2.
\end{align}
The 4-vertex region is thus given by the interval
\begin{align}
    \frac{\ell}{2} \leq a \leq \frac{\ell}{2} + | \alpha_2 |.
\end{align}
Note that it disappears when $\ell\rightarrow\infty$, which is consistent with this limit being the Witten vertex.

\subsection{Closed string Kaku vertices}
\label{section_closed_strings}

The closed string Kaku vertices are defined in terms of polyhedra on the Riemann surface, with the perimeter of the faces given by $L_i = |\alpha_i| + \ell$, and the bounds given by
\begin{align}
    L_\mathcal{I} \geq & ~ | \alpha_\mathcal{I} | + \ell.
\end{align}
As in the case of the open Kaku vertex, we can show that the reducible graphs never contribute. We again notice that the sum of the geodesics separating the $\mathcal{I}$ and $\mathcal{K}$ regions in the closed reducible graph in Figure \ref{reducible-graphs} is shorter than the perimeter of the $j$-th face. From there, the argument is identical to the open case. So here we will analyse only the irreducible graphs.

We observe that for $\ell \to \infty$ we get precisely the polyhedral vertices. To take this limit, we rescale the diagrams by $2\pi/\ell$. Then the external lengths are given by $L_i = 2\pi$, and the bounds on the closed curves become $L_\mathcal{I} \geq 2\pi$.

For $\ell=0$, we recover the lightcone vertices. In this case, we can easily show that all the $n$-vertices for $n\geq 4$ vanish, with a similar argument to the one used for the open string. First, note that, for the irreducible graphs, we always have three faces meeting at each vertex. Then, at least two of the faces that meet at any given vertex will have the same sign of momenta. Let's label these faces $i$ and $j$, and consider the closed path that encloses these two faces, of length $L_{ij}$. We have $L_{ij}\leq L_i+L_j$, since the two faces meet at at least one vertex. Since $\alpha_i$ and $\alpha_j$ have the same sign, we have
\begin{align}
    L_{ij} \leq |\alpha_i| + |\alpha_j| = |\alpha_i + \alpha_j|.
\end{align}
This, together with the geodesic bound on $L_{ij}$, implies $L_{ij} = |\alpha_i + \alpha_j|$, which only happens when the edge between the faces $i$ and $j$ has zero length (i.e., they only touch at the vertex). Thus, one parameter is fixed, and the vertex degenerates.

\begin{figure}[ht]
    \centering
    \includegraphics[width=.9\textwidth]{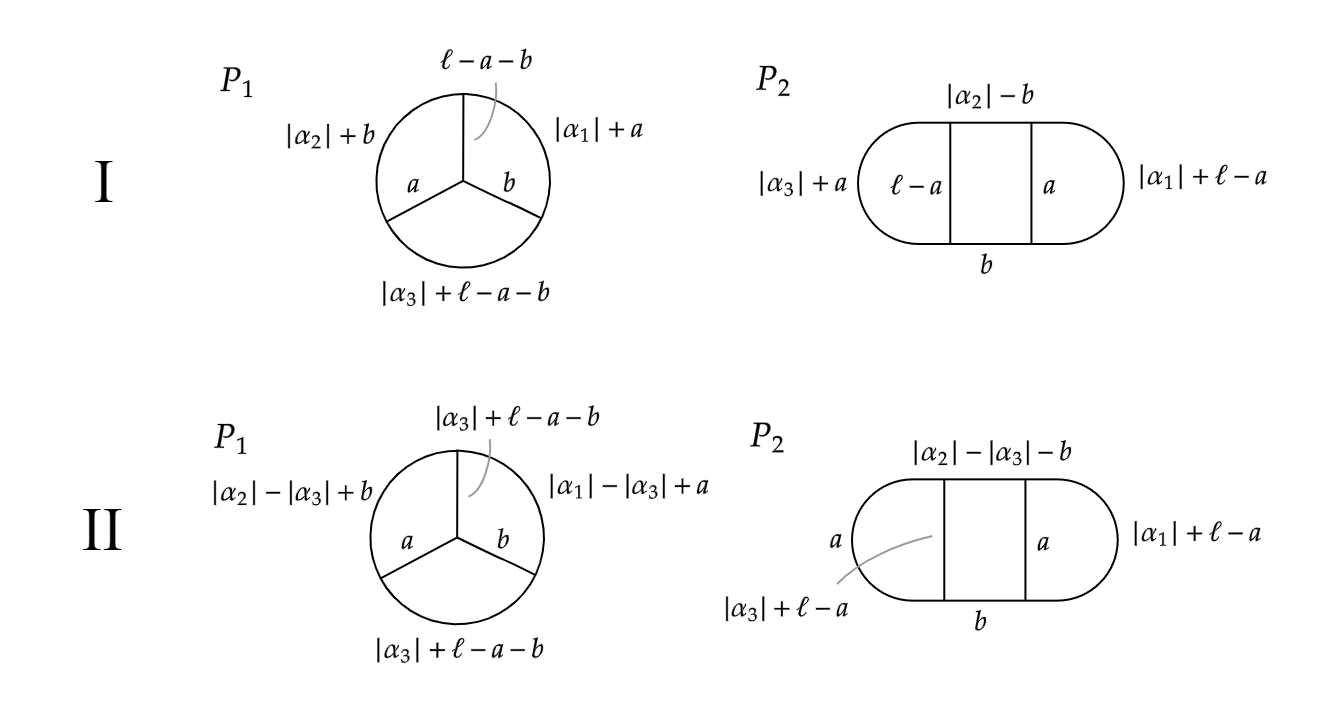}
    \caption{Parameterization of critical graphs, in Configurations I and II. We list only the irreducible graphs, which contribute to the 4-vertex.}
    \label{diagrams-closed-4-vertex}
\end{figure}

To characterize the geometry of the vertices for a fixed $\ell$, we will investigate how the 4-vertex $\mathcal{V}_4$ is contained in the moduli space $\mathcal{M}_4$. We consider the decomposition of $\mathcal{M}_4$ in patches defined by the critical graphs. A detailed analysis of this decomposition is given in Appendix \ref{appendix-closed-diagrams}. Here we will focus on analysing the irreducible graphs.

In the following, we will assume the inputs are in the {\it standard configuration}. It means that, for the $n$-vertex, the input parameter $\alpha_n$ has the biggest absolute value of all the inputs: $| \alpha_n | \geq | \alpha_i |$, $i=1,\cdots,n-1$, and its has negative sign: $\alpha_n \geq 0$.

We split the analysis in two input configurations. They are
\begin{align}
    \text{I. } & ~ |\alpha_1| + |\alpha_2| + |\alpha_3| = | \alpha_4 |, \nonumber \\
    \text{II. } & ~ |\alpha_1| + |\alpha_2| = | \alpha_3 | + | \alpha_4 |.
\end{align}
The graphs are parameterized differently in Configurations I and II. In Figure \ref{diagrams-closed-4-vertex} we show the parameterization of the irreducible graphs, which are the critical graphs that contribute to the 4-vertex.

\vspace{.1cm}
\noindent{\bf Configuration I.} The chart defined by $P_1$ has coordinates $(a,b)$, with
\begin{align}
    a \geq 0, \hspace{.6cm}
    b \geq 0, \hspace{.6cm}
    a+b \leq \ell.
\end{align}
The domain has the shape of a right triangle. To find the region of the 4-vertex in this patch, we look at the bounds on the lengths $L_{ij}$. We have
\begin{align}
    L_{14} = & ~ |\alpha_2| + |\alpha_3| + 2 \ell + 2a + 2b \geq |\alpha_2| + |\alpha_3| + \ell \implies a+b \geq \ell/2, \nonumber \\
    L_{24} = & ~ |\alpha_1| + |\alpha_3| + 2 \ell - 2a \geq |\alpha_1| + |\alpha_3| + \ell \implies a \leq \ell/2, \nonumber \\
    L_{34} = & ~ |\alpha_1| + |\alpha_2| + 2 \ell - 2 b \geq |\alpha_1| + |\alpha_2| + \ell \implies b \leq \ell/2.
\end{align}
This defines a triangle-shaped region, with orthogonal sides with length $\ell/2$. There is another $P_1$ patch obtained by switching $\alpha_1$ and $\alpha_2$, and one can check that it also contains part of the 4-vertex, which is again a triangle in these coordinates.

The domain of the coordinates $(a,b)$ for $P_2$ in Figure \ref{diagrams-closed-4-vertex} is
\begin{align}
    0 \leq a \leq \ell, \hspace{.6cm}
    0 \leq b \leq |\alpha_2|.
\end{align}
The bounds on $L_{14}$ and $L_{34}$ are
\begin{align}
    L_{14} = & ~ |\alpha_2| + |\alpha_3| + 2a \geq |\alpha_2| + |\alpha_3| + \ell \implies a \geq \ell/2, \nonumber \\
    L_{34} = & ~ |\alpha_1| + |\alpha_2| + 2\ell - 2a \geq |\alpha_1| + |\alpha_2| + \ell \implies a \leq \ell/2.
\end{align}
This fixes $a=\ell/2$. For the channel $\mathcal{I}=\{2,4\}$, there are two curves $L_{24}$ and $L_{24}'$, and we impose the bound on both:
\begin{align}
    L_{24} = & ~ |\alpha_1| + |\alpha_3| + 2\ell + b \geq | \alpha_1 | + | \alpha_3 | + \ell \implies b \geq - \ell, \nonumber \\
    L_{24}' = & ~ |\alpha_1| + |\alpha_2| + |\alpha_3| + 2\ell - b \geq | \alpha_1 | + |\alpha_3| + \ell \implies b \leq |\alpha_2| + \ell.
\end{align}
Neither of the bounds impose further constraints on $b$. The part of $\mathcal{V}_4$ contained in $P_2$ is just the curve $(a,b)=(\ell/2,b)$.

\begin{figure}[ht]
    \centering
    \includegraphics[width=\textwidth]{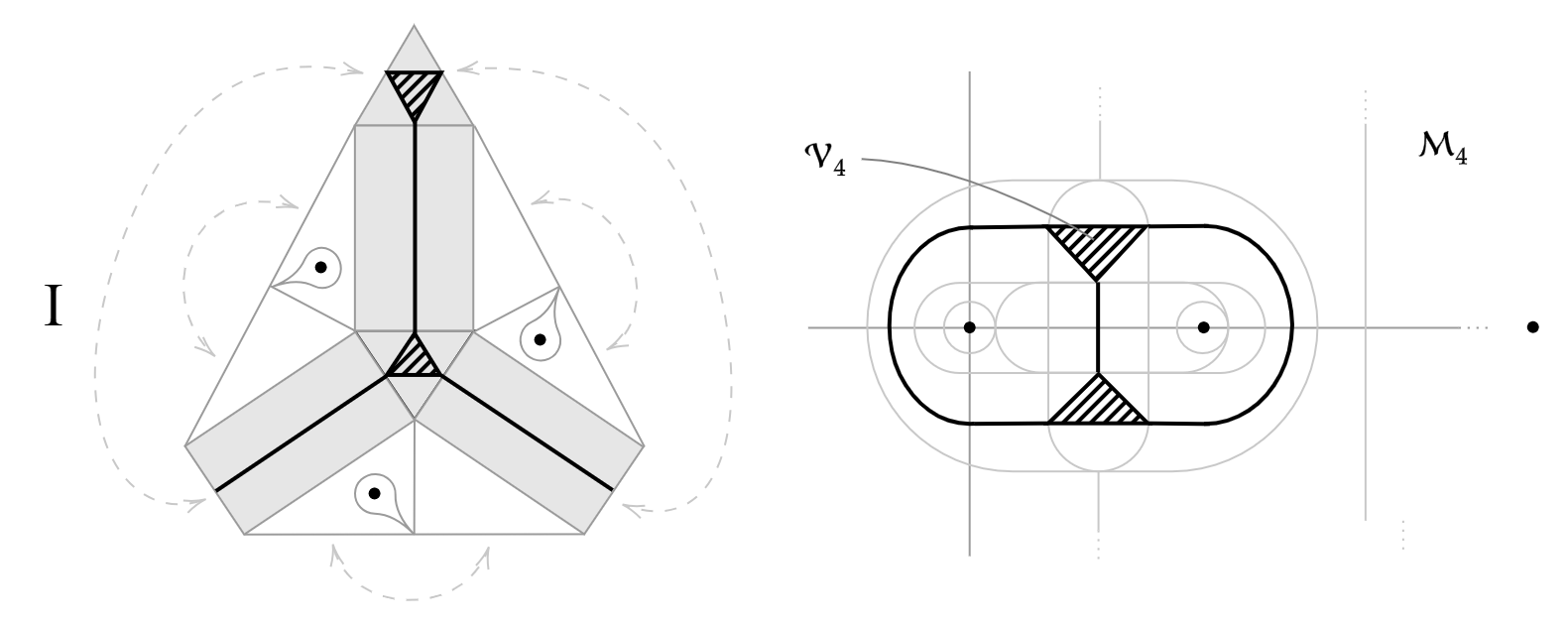}
    \caption{Visualization of the 4-vertex in the moduli space, in Configuration I. On the left, $\mathcal{M}_4$ is decomposed in patches corresponding to critical graphs. The patches shaded in grey correspond to irreducible graphs. On the right, $\mathcal{M}_4$ is represented schematically on the complex plane.}
    \label{moduli-space-3-1}
\end{figure}

There are two other configurations of $P_2$, obtained by making cyclic permutations of $(\alpha_1, \alpha_2, \alpha_3)$. They also contain the 4-vertex degenerated to a curve. Essentially, the 4-vertex vanishes in the $P_2$ patches, but it is important to keep track of these curves, as they show us how the non-degenerate parts of the 4-vertex connect.

The left diagram in Figure \ref{moduli-space-3-1} shows what is the geometry of $\mathcal{V}_4$ within $\mathcal{M}_4$, in Configuration I. We draw the decomposition of $\mathcal{M}_4$ in patches defined by the critical graphs. The patches shaded in grey correspond to the irreducible graphs, $P_1$ and $P_2$.

The boundaries in between the patches contain degenerate graphs: they are the ones obtained when the length of one of the edges goes to zero. Let's give an example of a critical graph in the boundary between $P_1$ and $P_2$ patches. From the parameterizations of Configuration I in Figure \ref{diagrams-closed-4-vertex}, consider the $P_1$ critical graph with parameters $(a,b)=(a,0)$. It is equivalent to the $P_2$ critical graph with parameters $(a,b)=(a,0)$ with $\alpha_1$ and $\alpha_3$ switched.

The 4-vertex region is contained in the irreducible patches. The other patches come from the reducible graphs, as explained in Appendix \ref{appendix-closed-diagrams}. The reducible graphs are labeled $P_3$ and $P_4$. They correspond respectively to the triangle and teardrop shaped patches. Their contribution to the decomposition of $\mathcal{M}_4$ is not important for the characterization of the 4-vertex, but we also draw their patches, for completeness.

The moduli space $\mathcal{M}_4$ has three singular points which correspond to degenerate Riemann surfaces. These points are also depicted on the left diagram in Figure \ref{moduli-space-3-1}, inside the $P_4$ patches.

The diagram on the right in Figure \ref{moduli-space-3-1} is another depiction of the geometry of $\mathcal{V}_4$ within $\mathcal{M}_4$. It is a schematic representation on the complex plane. The three singular points are fixed at $0$, $1$ and $\infty$. In the lightcone limit, $\ell\to 0$, the triangular regions disappear, and the 4-vertex is the joining of three lines which split the complex plane in three regions (which correspond to the $s-$, $t-$ and $u-$ channels). In the polyhedral limit, $\ell \to \infty$, the degenerate lines shrink and the two triangle shaped regions join their tips and form the known depiction of the polyhedral 4-vertex in the complex plane.

\vspace{.1cm}
\noindent{\bf Configuration II.} The patch $P_1$ has coordinates $(a,b)$ in the domain given by
\begin{align}
    a \geq 0, \hspace{.6cm}
    b \geq 0, \hspace{.6cm}
    a+b \leq |\alpha_3| + \ell,
\end{align}
which is a triangle with orthogonal sides of lengths $|\alpha_3|+\ell$. The bounds on the channel lengths are
\begin{align}
    L_{14} = & ~ - 2a + 2 \ell + \alpha_2 + | \alpha_3 | \geq \alpha_2 - | \alpha_3 | + \ell \implies a \leq | \alpha_3 | + \ell/2, \nonumber \\
    L_{24} = & ~ - 2b + 2\ell + \alpha_1 + | \alpha_3 | \geq \alpha_1 - | \alpha_3 | + \ell \implies b \leq | \alpha_3 | + \ell/2, \nonumber \\
    L_{34} = & ~ 2a + 2b + \alpha_1 + \alpha_2 - 2 | \alpha_3 | \geq \alpha_1 + \alpha_2 + \ell \implies a+b \geq | \alpha_3 | + \ell/2.
\end{align}
The 4-vertex region has the shape of a quadrilateral. In the other $P_1$ patch, obtained by switching $\alpha_1$ and $\alpha_2$, the 4-vertex region has the same geometry. The $P_1$ patches meet along one boundary, and the 4-vertex regions in each patch share one boundary.

There are also degenerate regions of the 4-vertex in the $P_2$ patches. For instance, in the $P_2$ diagram of Configuration II in Figure \ref{diagrams-closed-4-vertex}, the channel bounds set $(a,b) = (| \alpha_3 | +\ell/2, b)$, which is a one-dimensional region. There is another $P_2$ patch, obtained by switching $\alpha_1$ and $\alpha_2$, which contains another degenerate part of the 4-vertex.

\begin{figure}[ht]
    \centering
    \includegraphics[width=\textwidth]{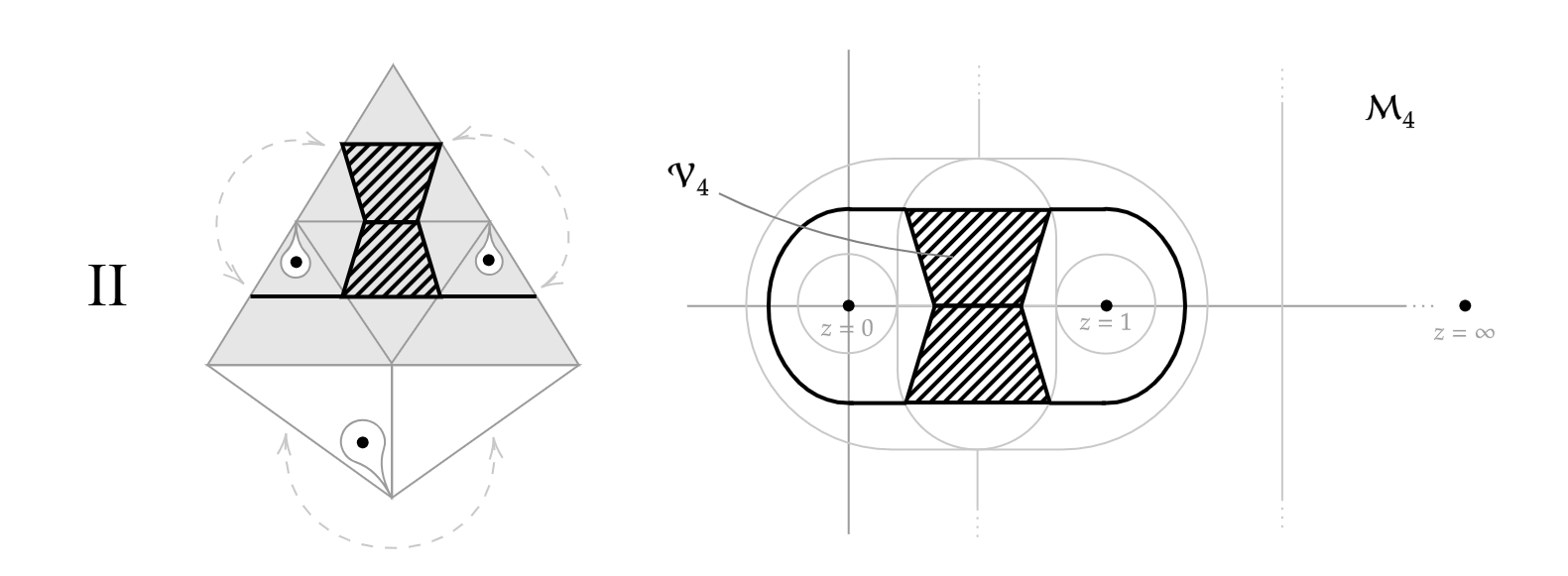}
    \caption{Visualization of the 4-vertex in the moduli space, in Configuration II. On the left, $\mathcal{M}_4$ is decomposed in patches corresponding to critical graphs. The patches shaded in grey correspond to irreducible graphs. On the right, $\mathcal{M}_4$ is represented schematically on the complex plane.}
    \label{moduli-space-2-2}
\end{figure}

The geometry of $\mathcal{V}_4$ is schematically shown in Figure \ref{moduli-space-2-2}. The diagram on the left shows the decomposition of $\mathcal{M}_4$ in patches of critical graphs. The patches shaded in grey correspond to irreducible diagrams. On the right we represent the 4-vertex on the complex plane. In the $\ell\to 0$ limit we obtaine the lightcone 4-vertex, which is degenerated to a line. In the $\ell\to \infty$ limit, we obtain the polyhedral 4-vertex. This limit is identical to the $\ell\to \infty$ limit in Configuration I, as the geometry becomes independent of the inputs $\alpha_i$.

\section{Conclusion}
\label{section_conclusion}

We have seen that it is possible to generalize the concept of hyperbolic vertices, if we allow the external lengths to be dependent on state inputs. The vertex condition on the systole of the surface is naturally generalized to geodesic bounds. Apart from containing the hyperbolic vertices within it, the two dimensional family of hyperbolic Kaku vertices contains the Kaku vertices in the flat limit of the hyperbolic geometry.

For open strings, the characterization of the flat limit in terms of critical graphs drawn on the open string Riemann surfaces gave us a nice and simple way of computing the vertices, and allowed us to check that the open string Kaku vertices are indeed obtained in the limit.

In the flat limit for closed strings we found the closed string Kaku vertices. The resulting one parameter family of vertices has the closed string lightcone vertex and the polyhedral vertices as limits, as expected. Thus we have characterized a family of vertices which interpolates between two important choices of vertices: the polyhedral vertices, which is a simple covariant choice of closed string vertices; and the closed string lightcone vertex, which is simple because it is cubic, even though it has no Lorentz covariance.

The definition of the hyperbolic Kaku vertices for higher genus remains as an open problem. The bounds are well defined for the hyperbolic vertices (with equal external lengths), but it is not clear how to define them when the external lengths are proportional to the momentum, since the loop momenta are not fixed by momentum conservation.

\section*{Acknowledgments}

We thank Ted Erler for suggesting the topic of this work, for useful discussions and for commenting on an earlier draft of the paper. We also thank Tomáš Procházka for useful discussions. UP thanks the Institute of Physics of the Czech Academy of Sciences for their hospitality during development of this work. The work of VB was supported by the Grant Agency of the Czech Republic under the grant EXPRO 20-25775X. The work of UP was supported by FAPESP grant number 2023/00862-4.

\appendix

\section{Critical graphs at $n=4$}
\label{appendix-critical-graphs}

Here we will describe what are the critical graphs used to parameterize the moduli space $\mathcal{M}_4$, for closed strings, and $\mathcal{M}_4^\text{disk}$, for open strings.

\subsection{Open strings}
\label{appendix-open-diagrams}

Describing the patch cover of $\mathcal{M}_4^\text{disk}$ for open strings is not necessary for the analysis of the open string Kaku vertices. However, it is helpful to understand the open string case before going to closed strings.

\begin{figure}[ht]
    \centering
    \includegraphics[width=.8\textwidth]{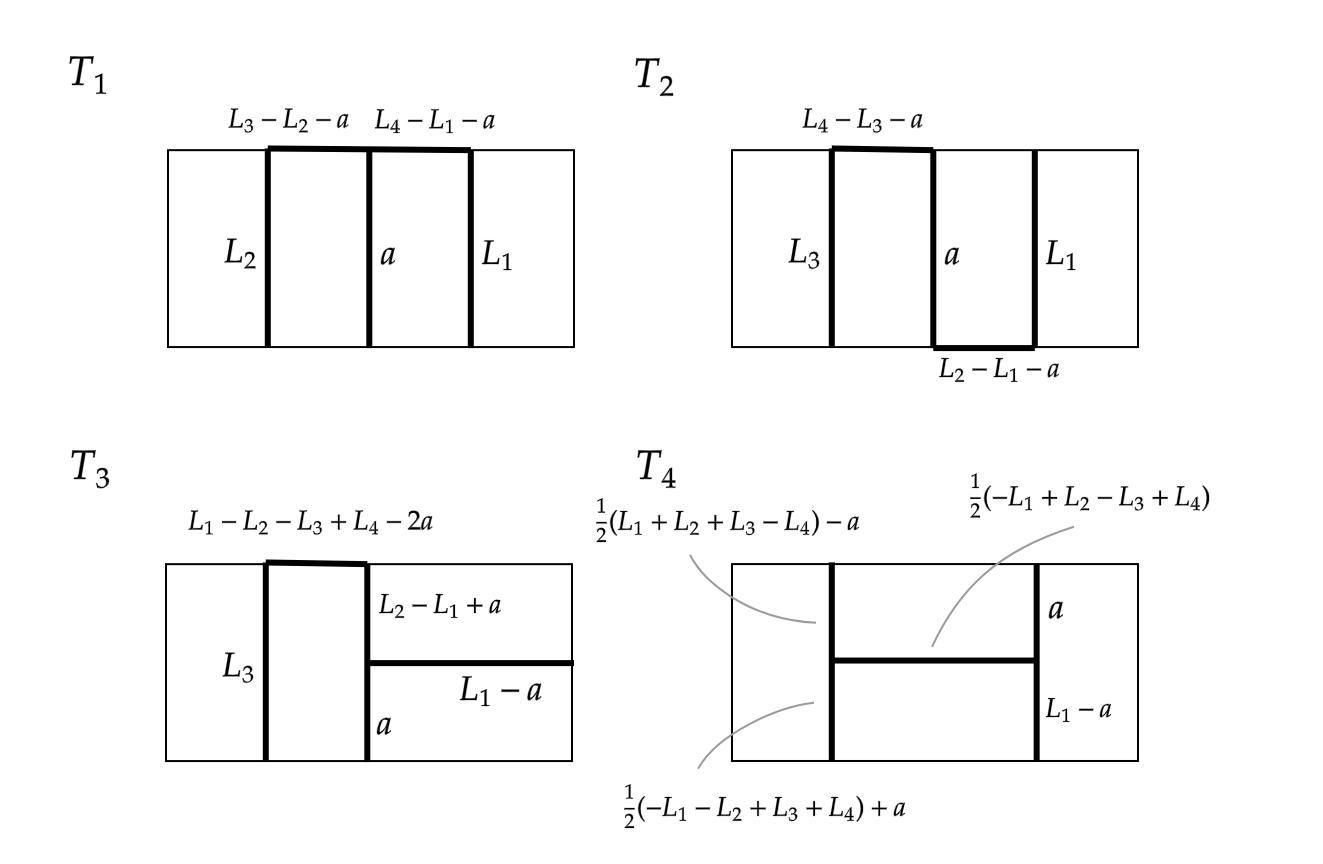}
    \caption{Critical graphs of the open string Riemann surface, for $n=4$.}
    \label{open-triangulations}
\end{figure}

Figure \ref{open-triangulations} shows the four critical graphs of the open string Riemann surface. We label them as $T_1$, $T_2$, $T_3$ and $T_4$. Each of them is a function of the external lengths $L_1$, $L_2$, $L_3$ and $L_4$. The thin lines represent the surface boundary, and the thick lines are edges of the diagram.

Each topology defines coordinates in a patch of the moduli space $\mathcal{M}_4^\text{disk}$. The patches never intersect, only share boundaries. For each patch, the corresponding ranges of the coordinate $a$ are:
\begin{align}
    T_1: \hspace{6.1cm}
    0 \leq a & ~ \leq \text{min} (L_3-L_2, L_4-L_1), \nonumber \\
    T_2: \hspace{6.1cm}
    0 \leq a & ~ \leq \text{min} (L_4-L_3, L_2-L_1), \nonumber \\
    T_3: \hspace{3.4cm}
    \text{max} (0, L_1-L_2) \leq a & ~ \leq \text{min} (L_1, \tfrac{1}{2} (L_1 - L_2 - L_3 + L_4) ), \nonumber \\
    T_4: \hspace{.6cm}
    \text{max} (0, \tfrac{1}{2} (-L_1-L_2+L_3+L_4)) \leq a & ~ \leq \text{min} (L_1, \tfrac{1}{2} (L_1+L_2+L_3-L_4)).
\end{align}

All of the graphs have edges along the boundary, except for $T_4$. Its geodesics $L_{12}$ and $L_{14}$ have the following lengths:
\begin{align}
    L_{12} = L_1 + L_2 - 2a, \hspace{.6cm}
    L_{14} = L_4 - L_1 + 2a.
\end{align}
For any choice of $\ell$ in the family of open string Kaku vertices, the 4-vertex is completely contained within the $T_4$ patch.

To study how the patches cover $\mathcal{M}_4^\text{disk}$, let's split the analysis in three configurations. We take $L_4 \geq L_i$, $i=1,2,3$, and we place the punctures in the order $L_1, L_2, L_3, L_4$. The three configurations are
\begin{align}
    \text{I. } & ~ |\alpha_1| + |\alpha_2| + |\alpha_3| = |\alpha_4|, \nonumber \\
    \text{IIa. } & ~ |\alpha_1| + |\alpha_2| = |\alpha_3| + |\alpha_4|, \nonumber \\
    \text{IIb. } & ~ |\alpha_1| + |\alpha_3| = |\alpha_2| + |\alpha_3|.
\end{align}
Configuration I is represented by a Mandelstam diagram of a string that splits in three. Configurations IIa and IIb are represented by Mandelstam diagrams of two strings rearranging again into two strings. We can refer to Configuration IIb as the non-planar one, because it is where the non-planar 4-vertex appears. It is characterized by alternating signs of the inputs (following the order along the boundary).

\begin{figure}[ht]
    \centering
    \includegraphics[width=.8\textwidth]{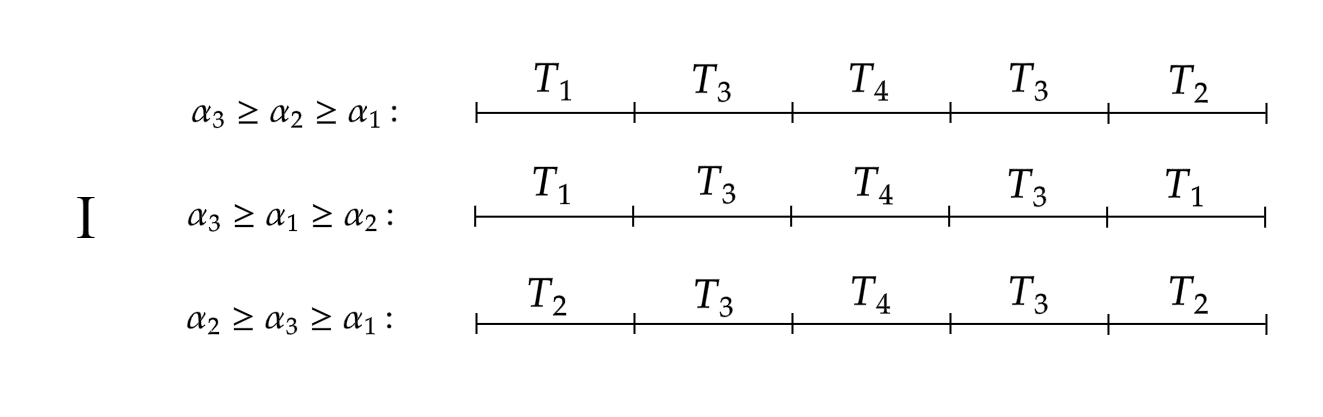}
    \caption{The cover of $\mathcal{M}_4^\text{disk}$ with open string critical graphs, in Configuration I. The corresponding critical graph is written above each patch.}
    \label{modulispace-open-3-1}
\end{figure}

The cover of $\mathcal{M}_4^\text{disk}$ with diagram patches in Configuration I is split in three cases. The decomposition of the moduli space in each case is depicted in Figure \ref{modulispace-open-3-1}. Different patches contribute to the cover depending on the absolute values of the inputs $\alpha_1$, $\alpha_2$ and $\alpha_3$. The length of the patch $T_4$ is $\ell$, so it vanishes in the lightcone limit ($\ell \to 0$), and it covers the entire moduli in the Witten limit ($\ell\to\infty$). The cases $\alpha_1 \geq \alpha_2 \geq \alpha_3$, $\alpha_1 \geq \alpha_3 \geq \alpha_2$ and $\alpha_2 \geq \alpha_1 \geq \alpha_3$ are obtained from the ones presented by a reflection operation (exchanging $\alpha_1$ and $\alpha_3$).

\begin{figure}[ht]
    \centering
    \includegraphics[width=.8\textwidth]{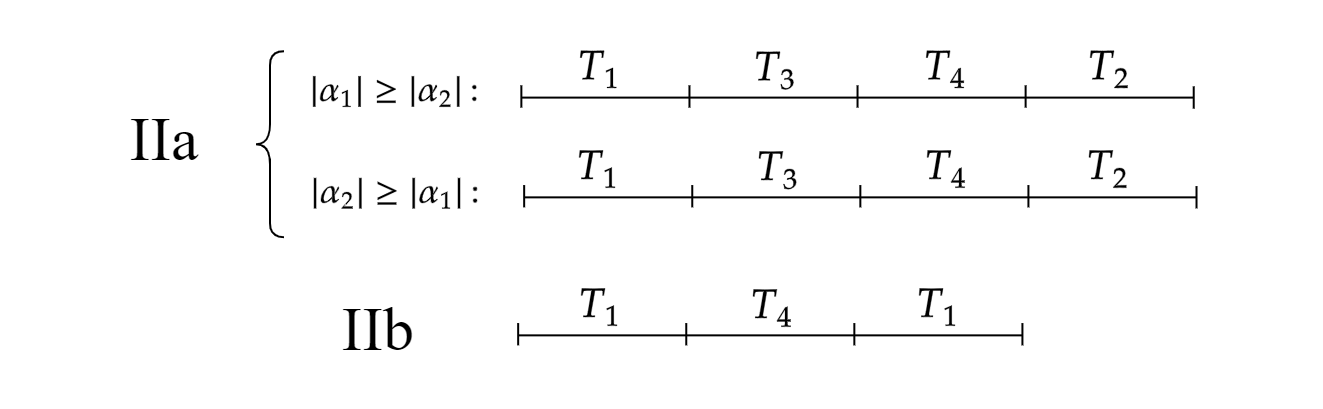}
    \caption{The cover of $\mathcal{M}_4^\text{disk}$ with open string critical graphs, in Configuration IIa, and Configuration IIb.}
    \label{modulispace-open-2-2}
\end{figure}

The cover of $\mathcal{M}_4^\text{disk}$ in Configuration IIa is split in two cases: $|\alpha_1| \geq |\alpha_2|$ and $|\alpha_2| \geq |\alpha_1|$. In this configuration, the length of $T_4$ is $|\alpha_3| + \ell$, so it doesn't vanish in the lightcone limit, but still it covers $\mathcal{M}_4^\text{disk}$ completely in the Witten limit. The configuration $|\alpha_2| + |\alpha_3| = |\alpha_1| + |\alpha_4|$ can be obtained from Configuration IIa by a reflection operation (so it exchanges $\alpha_1$ with $\alpha_3$). There is only one cover for Configuration IIb, and the length of $T_4$ is $| \alpha_2 | + \ell$, so the behaviour of this patch in the limits of $\ell$ is the same as in Configuration IIa. The patch covers for Configurations IIa and IIb are illustrated in Figure \ref{modulispace-open-2-2}.

\subsection{Closed strings}
\label{appendix-closed-diagrams}

\begin{figure}[ht]
    \centering
    \includegraphics[width=\textwidth]{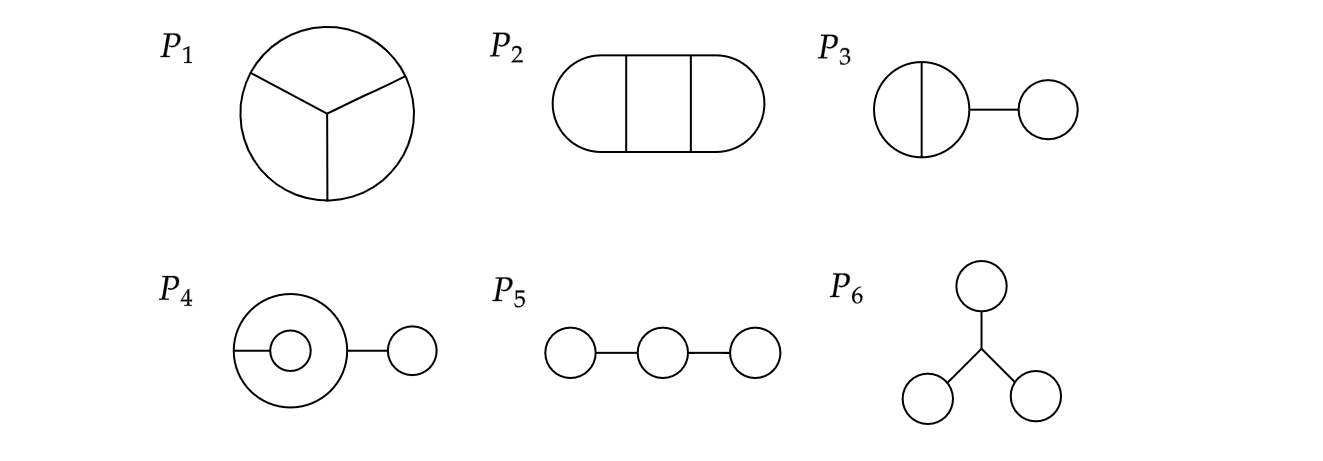}
    \caption{The 6 topologies of critical graphs with four faces on the closed Riemann surface. Each topology defines one type of tetrahedron.}
    \label{closed-diagrams}
\end{figure}

There are 6 different topologies of diagrams on closed Riemann surfaces, each of them defining a different polyhedron configuration. We label each diagram as $P_i$, for $i=1,\cdots, 6$. The letter $P$ stands for polyhedron. The diagrams are shown in Figure \ref{closed-diagrams}.

\begin{figure}[ht]
    \centering
    \includegraphics[width=.9\textwidth]{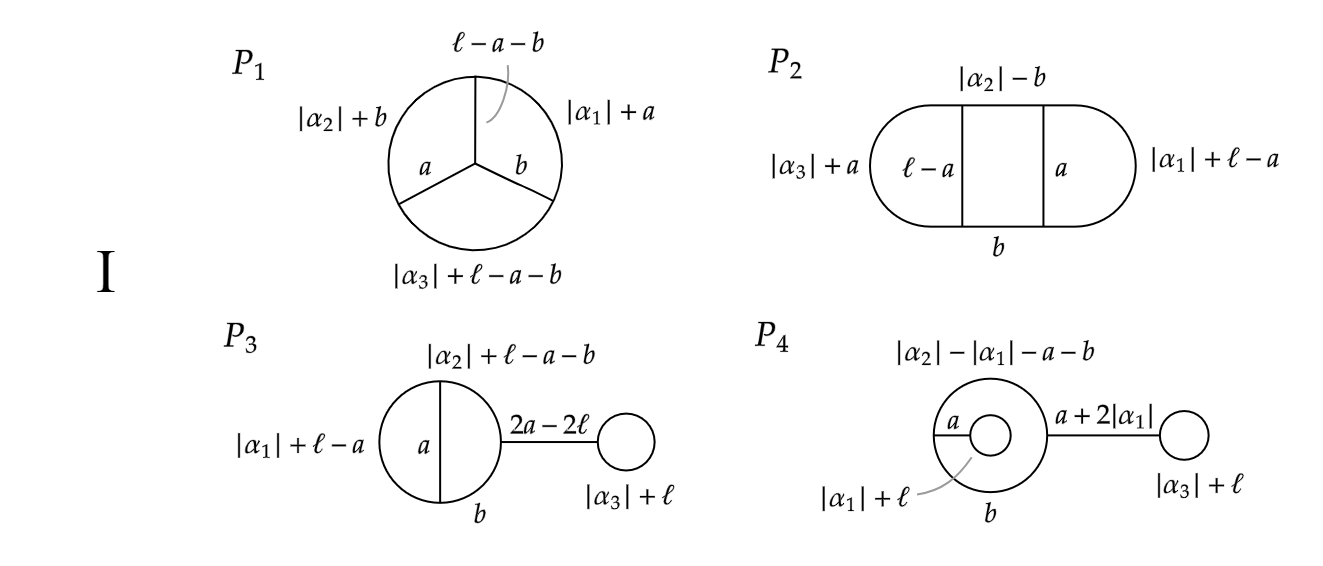}
    \caption{Parameterization of the critical graphs for the closed Kaku vertices, at Configuration I ($|\alpha_1|+|\alpha_2|+|\alpha_3|=|\alpha_4|$)}
    \label{closed-parameterization-3-1}
\end{figure}

For the closed Kaku vertices, we impose the constraints $L_i = |\alpha_i| + \ell$ on the perimeters of the faces of each diagram. There are no diagrams of topologies $P_5$ and $P_6$ that satisfy these constraints. To understand how the remaining diagram patches cover the moduli space $\mathcal{M}_4$, we split the analysis in two configurations
\begin{align}
    \text{I.} & ~ | \alpha_1 | + | \alpha_2 | + | \alpha_3 | = | \alpha_4 |, \nonumber \\
    \text{II.} & ~ | \alpha_1 | + | \alpha_2 | = | \alpha_3 | + | \alpha_4 |.
\end{align}
We will fix the face of perimeter $L_4 = |\alpha_4| + \ell$ to be the external face in all of the diagrams.

\vspace{.1cm}
\noindent{\bf Configuration I.} Figure \ref{closed-parameterization-3-1} shows the parameterization of the diagrams $P_i$, $i=1,\cdots,4$, in terms of coordinates $(a,b)$. Follows the description of the domain of $(a,b)$ in each patch:
\begin{itemize}
    \item For $P_1$, the domain of the coordinates $(a,b)$ is given by
\begin{align}
    a \geq 0, \hspace{.6cm}
    b \geq 0, \hspace{.6cm}
    a+b \leq \ell.
\end{align}
The chart is a triangle with sides proportional to $\ell$. As $\ell \to 0$, the chart vanishes. Each edge is a boundary with a $P_2$ patch.
    \item For $P_2$, the domain is
\begin{align}
    0 \leq a \leq \ell, \hspace{.6cm}
    0 \leq b \leq |\alpha_2|,
\end{align}
which is a chart of rectangular shape. As $\ell \to 0$, the patch vanishes. The boundary segments defined by extrema of $b$ are interfaces with $P_1$ patches, and the segments defined by the limiting values of $a$ are boundaries with $P_3$ patches.
    \item For $P_3$, the domain is
\begin{align}
    \ell \leq a \leq |\alpha_1| + \ell, \hspace{.6cm}
    b \geq 0, \hspace{.6cm}
    a+b \leq |\alpha_2| + \ell.
\end{align}
If $|\alpha_2| \leq |\alpha_1|$, this chart is triangular. The $a=\ell$ boundary is an interface with a $P_2$ patch. The other two boundaries connect the patch to a permuted $P_3$ patch (with the inputs $L_1$ and $L_2$ exchanged).

If $|\alpha_2| \geq |\alpha_1|$, the chart is rectangular, and the points $(a,b) = (|\alpha_1|+\ell, 0)$ and $(a,b) = (|\alpha_1|+\ell, |\alpha_2|-|\alpha_1|)$ are identified. So the chart is topologically a triangle with a hole that touches one of the vertices. The boundary at $a = \ell$ interfaces with a $P_2$ patch, and the boundaries at $b=0$ and $a+b = |\alpha_2|+\ell$ connect the patch with a permuted $P_3$ patch. And the $a=|\alpha_1|+\ell$ boundary, which is the boundary of the hole, connects with a $P_4$ patch.

    \item The $P_4$ patch (as drawn in Figure \ref{closed-parameterization-3-1}) only exists when $|\alpha_1| \leq |\alpha_2|$. The domain of the coordinates is given by
    \begin{align}
        a \geq 0, \hspace{.6cm}
        b \geq 0, \hspace{.6cm}
        a+b \leq |\alpha_2| - |\alpha_1|.
    \end{align}
    The boundaries at $b=0$ and at $a+b=|\alpha_2|-|\alpha_1|$ are identified, so the patch has the topology of a circle. The boundary at $a=0$ connects with a $P_3$ patch. In the point $(a,b)=(|\alpha_2|-|\alpha_1|,0)$ the diagram is not well-defined, and it corresponds to a singularity in the moduli space of diagrams. It is one of the three points in $\mathcal{M}_4$ where the Riemann surface degenerates.
\end{itemize}

\begin{figure}[ht]
    \centering
    \includegraphics[width=\textwidth]{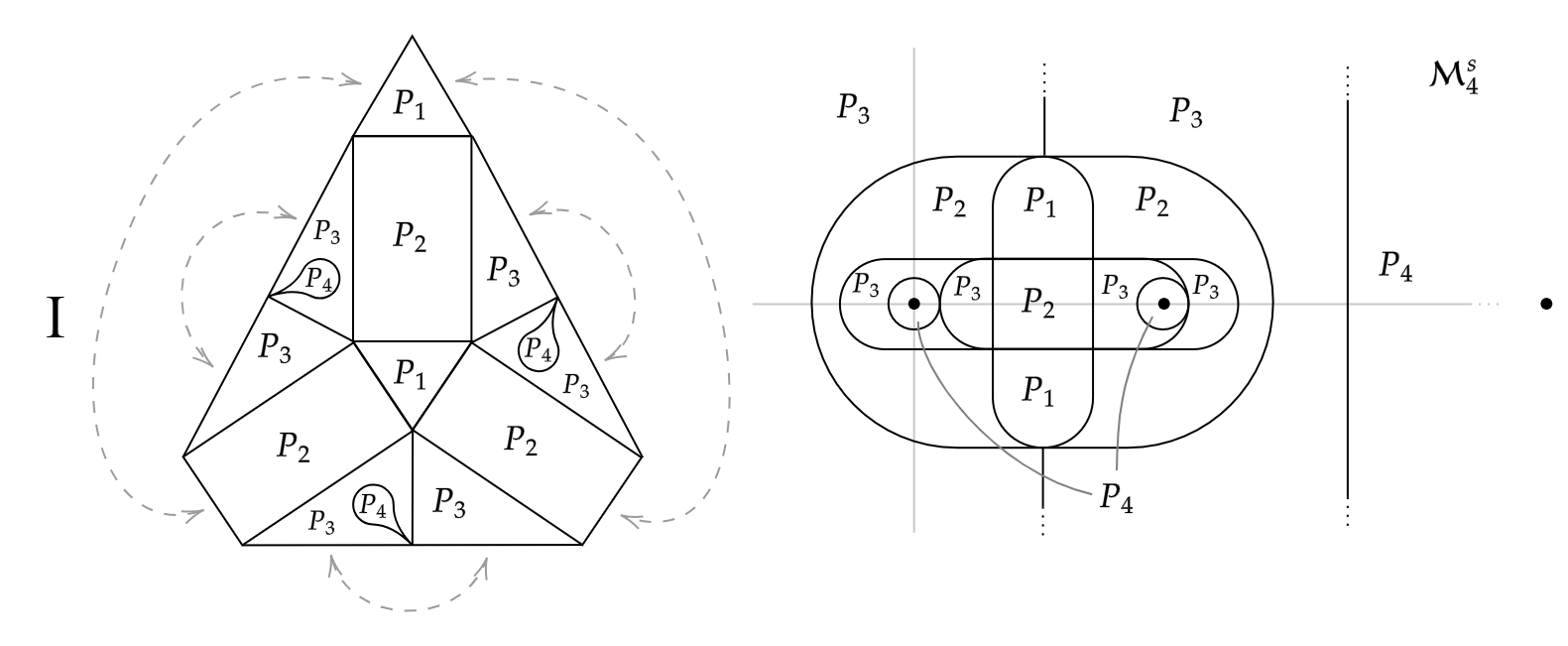}
    \caption{Two representations of the moduli space $\mathcal{M}_4$ covered with polyhedral patches patches in the Configuration I. The first is the visualization obtained from gluing the triangular and rectangular patches, and the second is a schematic visualization in the complex plane.}
    \label{modulispace-3-1}
\end{figure}

Figure \ref{modulispace-3-1} shows two representations of the moduli space $\mathcal{M}_4$ with polyhedral patches, in the Configuration I. One is the visualization of the gluing of each patch. The other one is a visualization in the complex plane, where the $z$ coordinate corresponds to the position of the fourth puncture. The fixed punctures are inside the $P_4$ patches.

\begin{figure}[ht]
    \centering
    \includegraphics[width=.9\textwidth]{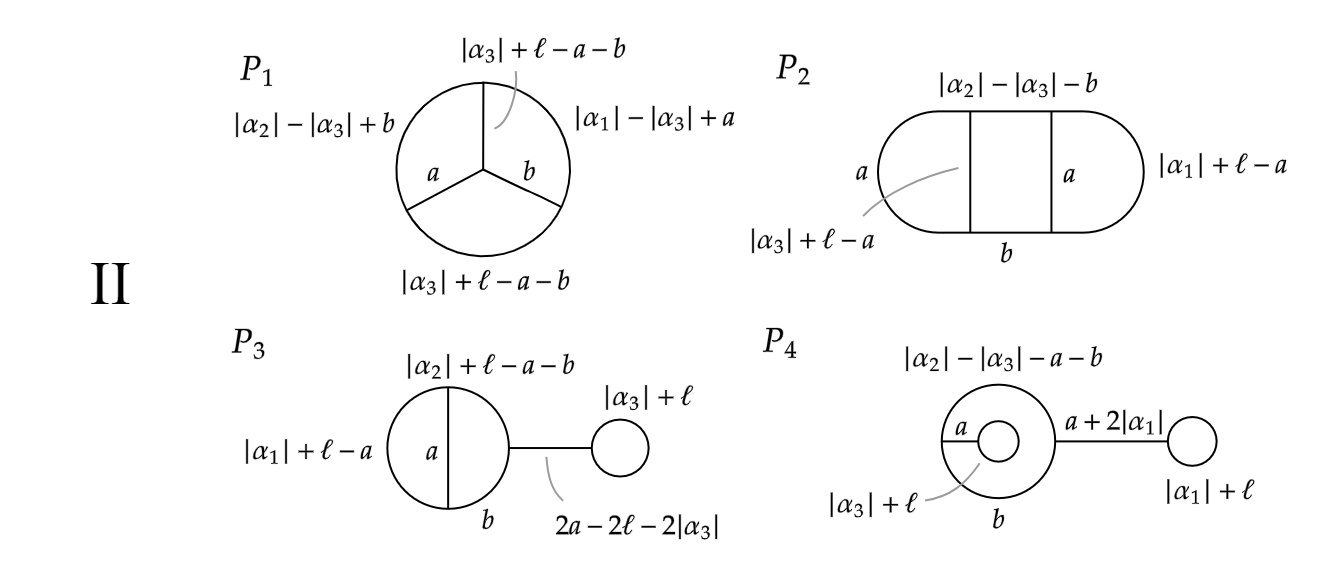}
    \caption{Parameterization of the critical graphs for the closed Kaku vertices, at Configuration II ($|\alpha_1|+|\alpha_2|=|\alpha_3|+|\alpha_4|$)}
    \label{closed-parameterization-2-2}
\end{figure}

\vspace{.1cm}
\noindent{\bf Configuration II.} The diagrams in this configuration are parameterized in a different way than in Configuration I. We recall that in the standard configuration we have $|\alpha_4| \geq |\alpha_1| \geq |\alpha_3|$ and $|\alpha_4| \geq |\alpha_2| \geq |\alpha_3|$. Figure \ref{closed-parameterization-2-2} shows the parameterization of the diagrams $P_i$, $i=1,\cdots,4$, in terms of coordinates $(a,b)$. Let's analyse the domains.
\begin{itemize}
    \item For $P_1$, the domain of the coordinates is given by
\begin{align}
    a \geq 0, \hspace{.6cm}
    b \geq 0, \hspace{.6cm}
    a+b \leq |\alpha_3| + \ell.
\end{align}
The chart is triangular. The boundaries at $a=0$ and $b=0$ are interfaces with $P_2$ patches, while the boundary at $a+b = |\alpha_3|+\ell$ connects the patch to the other $P_1$ patch (with $\alpha_1$ and $\alpha_2$ exchanged).
    \item For $P_2$, the domain is
\begin{align}
    0 \leq a \leq |\alpha_3| + \ell, \hspace{.6cm}
    0 \leq b \leq |\alpha_2| - |\alpha_2|,
\end{align}
which is a rectangular chart. The boundary segments defined by extrema of $b$ are interfaces with $P_1$ patches. The boundary at $a=|\alpha_3|+\ell$ is an interface with a $P_3$ patch. And the boundary segment at $a=0$ connects with a $P_4$ patch.
    \item For $P_3$, the domain is
\begin{align}
    |\alpha_3| + \ell \leq a \leq |\alpha_1| + \ell, \hspace{.6cm}
    b \geq 0, \hspace{.6cm}
    a+b \leq |\alpha_2| + \ell.
\end{align}
If $|\alpha_2| \leq |\alpha_1|$, this chart is triangular. The $a=|\alpha_3| + \ell$ boundary is an interface with a $P_2$ patch. The other two boundaries connect the patch to a permuted $P_3$ patch (with the inputs $L_1$ and $L_2$ exchanged).

If $|\alpha_2| \geq |\alpha_1|$, the chart is rectangular, and the points $(a,b) = (|\alpha_1|+\ell, 0)$ and $(a,b) = (|\alpha_1|+\ell, |\alpha_2|-|\alpha_1|)$ are identified. The chart is topologically a triangle with a hole that touches one of the vertices. The boundary at $a = |\alpha_3| + \ell$ interfaces with a $P_2$ patch, and the boundaries at $b=0$ and $a+b = |\alpha_2|+\ell$ connect the patch with a permuted $P_3$ patch. The $a=|\alpha_1|+\ell$ boundary, which is the boundary of the hole, connects with a $P_4$ patch.

    \item The $P_4$ patch, as drawn in Figure \ref{closed-parameterization-2-2}, has coordinates in the following domain
    \begin{align}
        a \geq 0, \hspace{.6cm}
        b \geq 0, \hspace{.6cm}
        a+b \leq |\alpha_2| - |\alpha_3|.
    \end{align}
    The boundaries at $b=0$ and at $a+b=|\alpha_2|-|\alpha_3|$ are identified, so the patch has the topology of a circle. The boundary at $a=0$ connects with a $P_3$ patch. In the point $(a,b)=(|\alpha_2|-|\alpha_3|,0)$ the diagram is not well-defined, and it corresponds to a singularity in $\mathcal{M}_4$.
\end{itemize}

\begin{figure}[ht]
    \centering
    \includegraphics[width=\textwidth]{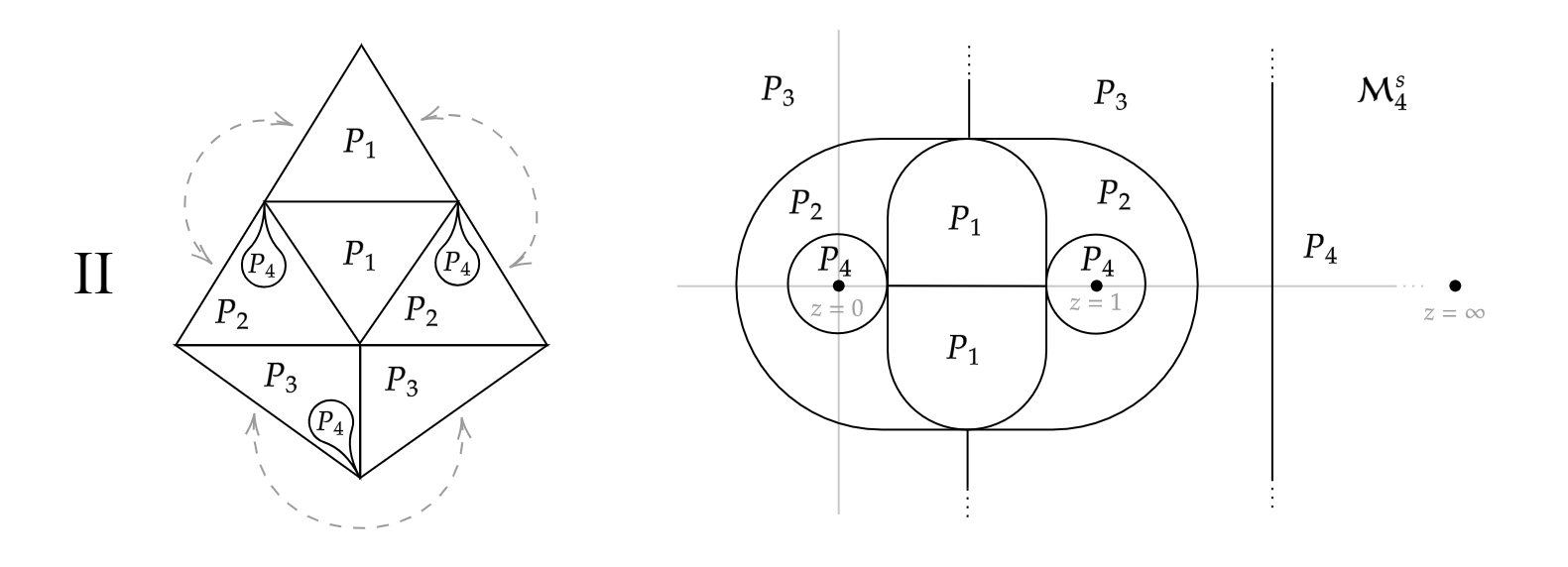}
    \caption{Two representations of the moduli space $\mathcal{M}_4$ covered with polyhedral patches in the Configuration II. The first is the visualization obtained from gluing the triangular and rectangular patches, and the second is a schematic visualization in the complex plane.}
    \label{modulispace-2-2}
\end{figure}

\bibliographystyle{hieeetr.bst}
\bibliography{refs}

\end{document}